\documentclass[12pt]{article}

\usepackage{epsfig}
\usepackage{amsmath,amssymb,amsfonts}
\usepackage{float,a4wide}
\usepackage{hyperref}
\newcommand{\be}{\begin{equation}}
\newcommand{\en}{\end{equation}}
\newcommand{\bea}{\begin{eqnarray}}
\newcommand{\ena}{\end{eqnarray}}

\newcommand{\hbo}{\hbox to 1 true cm {\hfill } }
\newcommand{\tr}{\hbox{tr}}

\def\dslash{\partial\kern-.5em\slash}
\def\kslash{k\kern-.5em\slash}
\def\pslash{p\kern-.5em\slash}
\def\Dslash{D\kern-.5em\slash}

\begin{document}
\vbox{
\hfill December 6, 2010
}

\vfil
\centerline{\large\bf Confinement from semiclassical gluon fields in SU(2)
gauge theory}

\bigskip\bigskip
\bigskip\bigskip
\centerline{Kurt Langfeld }
\vspace{.5 true cm}
\centerline{ School of Computing \& Statistics, University of Plymouth }
\centerline{Plymouth, PL4 8AA, UK }

\bigskip\bigskip
\centerline{and}

\bigskip\bigskip
\centerline{Ernst-Michael Ilgenfritz }
\vspace{.5 true cm}
\centerline{Institut f\"ur Physik, Humboldt-Universit\"at zu Berlin }
\centerline{D-12489 Berlin, Germany }

\vskip 1.5cm

\begin{abstract}
The infrared structure of $SU(2)$ Yang-Mills theory is studied
by means of lattice gauge simulations using a new {\it constrained
cooling} technique. This method reduces the action while
all Polyakov lines on the lattice remain unchanged. 
In contrast to unconstrained cooling, quark confinement is still intact. 
A study of the Hessian of the Yang-Mills action
shows that low action (semi-) classical configurations can be achieved,
with a characteristic splitting between collective modes and higher 
momentum modes. Besides confinement, the semiclassical configurations also
support the  topological susceptibility and generate spontaneous breakdown of
chiral symmetry.We show that they possess a cluster structure of locally
mainly (anti-) selfdual objects. By contrast to an instanton or a meron
medium, the topological charge of individual clusters is smoothly distributed.
\end{abstract}

\vfil
\hrule width 5truecm
\vskip .2truecm
\begin{quote}
\end{quote}
\eject

\section{Introduction}
\label{sec:Introduction}

More than thirty years ago, Callan, Dashen and
Gross~\cite{Callan:1977gz,Callan:1978bm} (CDG) attempted to derive the
non-perturbative properties of QCD from 
classical solutions of the Euclidean equations of motion in Yang-Mills theory. 
Such solution had been found few years before and called pseudoparticles or 
instantons~\cite{Belavin:1975fg}, carrying unit winding number (topological
charge). The program initiated by CDG was only partly successful: while chiral 
symmetry breaking became immediately understandable, confinement remained 
unexplained in terms of instantons. Thus, CDG assigned the property of
confinement  
to singular solutions (merons), objects with infinite
action~\cite{Callan:1977qs}  
and half-integer topological charge. The phase of (confining) QCD was thought to 
be realised on top of a Kosterlitz-Thouless transition where the dissociation 
of instantons into merons is probed by a action/entropy trade-off. This mechanism, 
however, was beyond the possibilities of a semiclassical calculation and was not 
worked out further at that time.

\vskip 0.3cm
Since lattice simulations have become possible, one of the aims was to
learn about the mechanism of confinement from the structure of lattice
configurations. The simplest method to search for a non-perturbative background
is cooling~\cite{Teper:1985ek,Ilgenfritz:1985dz}. Normal cooling makes visible
a granular structure of action and topological charge. It appeared natural to
interprete the lumps as instantons. Using improved actions, even unstable
instanton-antiinstanton pairs became relatively stabilised and observable under
cooling~\cite{de Forcrand:1997sq}.

\vskip 0.3cm
The problem is when cooling should be stopped. Ignoring ad-hoc recipes
derived from the instanton model itself, self-stopping algorithms are preferable.
Restricted cooling~\cite{GarciaPerez:1998ru,GarciaPerez:1999it} is such an example.
It tolerates a certain violation of the classical lattice equation of motions
in order to locally stop the relaxation. It turns out that this imposes a certain 
scale (other than the lattice spacing) and leads to a scale-dependent 
characterisation of the vacuum structure. Topological structure with a clear
scale dependence is also provided by spectral filtering techniques based on 
the lattice Dirac or Laplacian 
operator~\cite{Horvath:2002yn,Ilgenfritz:2007xu,Bruckmann:2005hy,Ilgenfritz:2008ia},
respectively, which do not modify the underlying gauge field.

\vskip 0.3cm
So far, the most diligent approach making use of cooling was a combination of 
blocking and inverse blocking. Here, inverse blocking meant cooling on a fine 
lattice with a perfect action, constrained to preserve the long-range structure 
kept in form of blocked lattice configurations~\cite{DeGrand:1996ih,DeGrand:1996zb,Feurstein:1996cf,Feurstein:1997rj}.
In this form, iterated cycles of blocking and inverse blocking avoid the collapse
to the trivial vacuum. It automatically finishes with a nontrivial ``background''
configuration stabilised by the blocked configuration. This numerically very 
expensive procedure was later mapped to simple APE-type smearing~\cite{DeGrand:1997gu}.
By extrapolating back to the ``no smearing'' limit, the instanton content of 
$SU(2)$~\cite{DeGrand:1997ss} and $SU(3)$~\cite{Hasenfratz:1998qk} pure gauge 
theory and full QCD with dynamical fermions~\cite{Hasenfratz:1999ng} has been 
described.

\vskip 0.3cm
There is a caveat, however. The attempted reconstruction of the lattice
configurations by reduction to their ``instanton content'' (positions and sizes)
does not reproduce confinement and the correct hadron spectrum~\cite{DeGrand:1997sd}.
Kovacs~\cite{Kovacs:1999an} showed that other (low- action) degrees of freedom
(torons) must be extracted from the individual lattice configurations as well
and added to the model configurations in order to mimic the original lattice
ensemble in a satisfactory way.

\vskip 0.3cm
At finite temperature, calorons (finite-$T$ instanton solutions) more general
than previously known have been found and described by Kraan and van
Baal~\cite{Kraan:1998pm,Kraan:1998sn} and Lee and Lu~\cite{Lee:1998bb}). These
solutions have two advantages over instantons: they are explicitely dependent 
on the embedding holonomy, and the moduli space contains the necessary degrees
of freedom to permit dissociation into $N_c$ constituents, dyons carrying 
fractional topological charge. One of the dyons carries the zero mode that would 
otherwise assigned to the whole caloron. A gas of moderately dissociated
calorons is able to predict confinement {\it provided} the holonomy parameter
is maximally nontrivial (i.e. the external Polyakov loop 
vanishes)~\cite{Gerhold:2006sk,Bruckmann:2009pa}.
Although this semiclassical model has not been completely worked out to the 
extent that the 
effective potential of the Polyakov loop {\it and} the confining force could be 
obtained as functions of the temperature, the expected caloron/dyon structures 
could be actually detected on realistic lattice configurations below
the deconfinement 
transition~\cite{Ilgenfritz:2002qs,Ilgenfritz:2006ju,Bornyakov:2007fm}.
In later studies cooling has not been used. Instead, modes of chiral (overlap) 
fermions were used as probes. In the deconfined phase the symmetry between dyons 
is broken, and only the light ones are abundant~\cite{Bornyakov:2008im}.

\vskip 0.3cm
In a model-independent way, the question of confinement facilitated by 
semiclassical configurations has been asked again and affirmatively answered by 
Gonzalez Arroyo and collaborators
~\cite{GonzalezArroyo:1995ex,GonzalezArroyo:1996jp,GonzalezArroyo:1996gs}.
They have used cooling, however employing twisted boundary 
conditions~\cite{GarciaPerez:1989gt,GarciaPerez:1992fj,GarciaPerez:1993ab}.
Under these circumstances a trivial vacuum is avoided and fractional topological 
charges can be observed.

\vskip 0.3cm
At present, a semiclassical picture for confinement seems not very popular. 
As far as a mechanism 
of confinement is concerned, the overwhelming opinion is that exclusively defects 
like vortices and monopoles~\cite{Greensite:2003bk,Alkofer:2006fu} play this 
role~\footnote{We refer to Ref.~\cite{Zakharov:2006vt} for a way to 
reconcile the 
two points of view.}. Practically, they are exposed as defects of gauge fixing. 
It is not excluded, however, that certain non-Abelian extended objects
(e.g. ``thick vortices''~\cite{GonzalezArroyo:1998ez}) exist, which become
merely {\it localised} by the process of gauge fixing and projection to the
Abelian degrees of freedom. We should mention that also calorons lead to a 
characteristic monopole and vortex structure~\cite{Bruckmann:2009pa}.

\vskip 0.3cm
A common feature of the last examples, beginning with inverse blocking, is the
interplay of topological configurations and certain constraints or boundary
conditions imposed to the cooling process. We think it is worth to pursue this 
guiding idea further. In a preceding paper, one of us has explored the 
feasibility of constrained cooling that preserves the property of 
confinement~\cite{Langfeld:2009es} by keeping fixed Wilson loop ratios in the 
given configuration at a given ``macroscopic'' (infrared) length scale. 
The cooling constrained by large-scale (infrared) observables is formulated
by adding Lagrange multiplier forces to the lattice equations of motion. 
This constrained cooling has also led to a gas formed by clusters of action. 
Not surprisingly, the cooled configurations reproduce the conventional
confinement criteria despite the low action and the apparent absence of 
strong fluctuations.  

\vskip 0.3cm
In this paper we explore the possibilities of another type 
of constrained cooling. It uses the information stored in the (locally varying) 
Polyakov loop. Because we study first the situation at $T=0$, we preserve the 
Polyakov loops in all (temporal and spatial) directions.

\vskip 0.3cm
The paper is organised as follows.
In the following Sect. \ref{sec:Streamlines} we consider
in general terms the connection between classical configurations, constrained
cooling and streamline configurations. In Sect. \ref{sec:ConstrainedCooling} 
we explain how cooling constrained by Polyakov loops can be realised.
Here we report about the easiest available observations, the behaviour of
action and Creutz ratios under cooling, and contrast this with standard
unconstrained cooling: confinement is preserved.
The next Sect. \ref{sec:GluonicSpectra} is devoted to the spectrum of
gluonic excitation modes on the background of streamline configurations and
the difference to the spectrum on standard-cooled configurations.
In Sect. \ref{sec:SpaceTimeTexture} we describe the lumpy
structure in terms of action and topological charge. In addition, the
degree of (anti-)selfduality and the localisation and dimensionality of
action and topological density is analysed. Sect. \ref{sec:SSB}
describes how the spectrum of the staggered Dirac operator depends on
constrained cooling and gives a comparison with standard cooling. In the 
case of constrained cooling we find that a set on near-zero modes survives, 
indicating that chiral symmetry breaking survives. In Sect.~\ref{sec:conclusions} 
we draw conclusions for further work. In an Appendix more details about solving 
the constraints during cooling are worked out.

\section{ Confinement from Yang-Mills streamlines  }

\subsection{Classical configurations, streamlines and cooling }
\label{sec:Streamlines}

As an illustration, let us consider a scalar field $\phi (x)$ with the Euclidean 
action $S[\phi]$.
Let the action possess several degenerate absolute minima which might
arise from a discrete symmetry of the action. These minima are separated by
potential wells over the multidimensional configuration space. The well
known double-well problem~\cite{Huang:1996wj} might serve as an example:
there, the two  minima of the action are related by the reflection $\phi \to -
\phi $ of  the corresponding scalar fields. If this symmetry is not
spontaneously (or anomalously) broken, a semi-classical calculation of the
partition function starts by taking into account the contributions
of all absolute minima weighted by the contribution from Gaussian
fluctuations around each ground state configuration. The implicit
assumption for a semi-classical treatment is that only configurations
close to these absolute vacuum states are relevant and tunnelling is
rare.
It is evident that this treatment breaks down if the
potential wells between the minima are small and tunnelling 
becomes frequent. Formally, this can be seen that space-time dependent
field configurations (e.g.~kinks) which interpolate in configuration space
between two possible vacuum states possess an action close to the
vacuum action implying that, depending on entropy, those kink
configurations are relevant as well in the full quantum theory.

\vskip 0.3cm
Low action solutions are generically obtained by the so-called
cooling procedures which minimises the action. If we consider a gas
of as many kink as anti-kinks, excessive cooling results in
an annihilation of kink anti-kink pairs leading to the trivial
vacuum. If interesting non-perturbative phenomenons can be attributed
to the kink configurations, cooling must be stopped by either an ad
hoc criterion or by additional physical input.

\vskip 0.3cm
Reducing the action by means of the steepest descent method, 
in other words
{\it standard cooling}, can be considered as a flow in configuration space
where the field picks up an additional dependence on the ``cooling
time'' $\tau$, i.e., $\phi (x) \to \phi (\tau , x) $.
By {\it constrained cooling}, we address a method which reduces the
action while preserving a constraint, such as
\be
\sum_x \; n(x) \, \phi (x) \; = \; \hbox{constant}, \hbo
\sum_x \;  n^2(x) \; = \; 1 \; .
\label{eq:ir0}
\en
Both cooling methods can be formally defined by the flow equation:
\be
\frac{ \partial \phi(\tau,x) }{ \partial \tau } \; = \;
- \, \frac{ \delta S[\phi] }{\delta \phi(\tau,x) } \; + \;
\left( \sum _y \;  n(y) \; \frac{ \delta S[\phi] }{\delta \phi(\tau,y) }
\right) \, n(x) \; .
\label{eq:ir1}
\en
It is easy to show that
$$
\frac{ \partial }{\partial \tau } \, \sum_x \; n(x) \,
\phi(\tau,x) \; = \; 0
$$
implying that the constraint (\ref{eq:ir0}) has been implemented exactly.
Note that the flow equation for standard cooling is obtained by
setting $n(x) = 0$.

\vskip 0.3cm
Consider $\phi^c(x)$ as a low action solution for the case the constraint
has been switched off, 
i.e., $n(x)=0$. 
If $\phi^c(x)$ is, according to
some norm, close to a minimum configuration $\phi _\mathrm{min}(x)$,
the difference between $\phi ^c(x)$ and $\phi (\tau,x)$,
$$
\Delta \phi (\tau, x) \; := \; \phi(\tau, x) - \phi^c(x) \; ,
$$
is small for some large value $\tau $. Note that, {\it with}
the constraint in place, the
constraint has to be sufficiently weak such that  $\Delta \phi$ can
be small. Whether this is indeed the case for a particular theory
has to be justified by a numerical experiment.
With the above assumption, we might then approximate the action by a second
order polynomial:
\bea
S[\phi] &=& S[\phi^c] \; + \; \sum_x \, \Delta \phi (x) \; f[\phi^c](x)
\label{eq:ir2} \\
&+& \frac{1}{2} \sum_{xy} \; \Delta \phi (x) \; M[\phi^c](x,y) \;
\Delta \phi (y) \; + \; {\cal O}(\Delta \phi ^3) \; .
\nonumber
\ena
The drift force $f$ and Hessian $M$ are thereby given by
\be
f[\phi^c](x) \; = \; \frac{\delta S[\phi]}{\delta \phi(x) } \Bigr\vert
_{\phi=\phi^c}, 
\hbo
M[\phi^c](x,y) \; = \; \frac{\delta^2 S[\phi]}{\delta \phi(x) \, \delta
  \phi(x) }
\Bigr\vert _{\phi=\phi^c}.
\label{eq:ir3}
\en
Inserting (\ref{eq:ir2}) in the flow equation (\ref{eq:ir1}), we find:
\be
\frac{ \partial \phi (\tau,x) }{\partial \tau } \; = \;  - \, f_x \, - \,
M_{xy} \, \Delta \phi _y \; + \; \Bigl( n_y [
f_y + M_{yz} \Delta \phi_z] \Bigr) \, n_x \; ,
\label{eq:ir4}
\en
where we have introduced the convention to sum over indices which
appear twice. Introducing the projector
$$
P_{xy} \; = \; \delta _{xy} \, - \, n_x \, n_y \; , \hbo P^2 = P \, ,
$$
the last equation can be written in compact notation:
\bea
\frac{ \partial \phi (\tau,x) }{\partial \tau } &=&   - \, F_x \, - \,
{\cal M}_{xy} \, \Delta \phi _y \; ,
\label{eq:ir5} \\
F&=&  P \, f \; , \hbo
{\cal M} \; = \; P \, M  \; .
\label{eq:ir6}
\ena
The flow equation can be easily solved
\be
\phi (\tau ) \; = \; \phi ^c \; + \; \exp \Bigl\{ -  {\cal M} \, \tau
\Bigr\} \, \phi(0) \; - \; \Bigl[ 1 - \exp \Bigl\{ -  {\cal M} \, \tau
\Bigr\}  \Bigr] \; G \; , 
\label{eq:ir7}
\en
where
$$
{\cal M} G \; = \; F \, \hbo \Rightarrow \hbo G \; = \; (PMP)^{-1} \, F
$$
given that $PMP$ is invertible in the projected space.

\vskip 0.3cm
We firstly study {\it standard cooling} where we have:
$$
n(x)=0, \hbo {\cal M}=M=M^T, \hbo F=f \; .
$$
Since the Hessian $M$ is symmetric, we can introduce its eigenmodes
as a complete basis, i.e.,
\be
M_{xy} \; \psi^{(n)}_y \; = \; \lambda _n \; \psi ^{(n)}_x \; , \hbo
\lambda _1 \le \lambda _2 \le \ldots
\label{eq:ir8}
\en
and expand
\be
\Delta \phi _x \; = \; \sum_n c_n(\tau ) \; \psi ^{(n)}_x \; , \hbo
f[\phi^c]_x \; = \; \sum _n f_n \, \psi^{(n)}_x \; .
\label{eq:ir9}
\en
The solution (\ref{eq:ir7}) of the flow equation now translates into
\be
c_n(\tau) = c_n(0) \, \exp \{- \lambda _n \tau \} \; - \;
\frac{ f_n }{\lambda _n} \Bigl[ 1 - \exp \{- \lambda _n \tau \} \Bigr] \; .
\label{eq:ir10}
\en
In the limit $\tau \to \infty$, we observe $c_n \to - f_n/\lambda _n $ and
$\phi $ approaches the true minimum of the action (\ref{eq:ir2}),
i.e., $\phi \to \phi _\mathrm{min} (x)$:
\be
\lim _{\tau \to \infty} \phi (\tau) \; = \; \phi ^c \; - \;
M^{-1} \, f \; = \; \phi _\mathrm{min} \; .
\label{eq:ir10b}
\en
The crucial idea is to
{\it stop cooling} by only admitting $\tau \le \tau _\mathrm{stop}$.
As explained by Huang in~\cite{Huang:1996wj}, the success of a
semi-classical expansion critically depends on the existence of a gap
in the eigenmode spectrum of the Hessian. Assume that such a gap
occurs at the $i$th eigenvalue, i.e., $\lambda _i \ll \lambda _{i+1}$.
In this case, we will find a range $R$ of $\tau $ values with
\be
R \; = \; \Bigl\{ \tau \in [\tau_l, \tau _h] \; \Big\vert \;
\lambda _i \tau_h \ll 1, \; \lambda _{i+1} \tau _l \gg 1 \Bigr\} \; .
\label{eq:ir11}
\en
For $\tau \in R$, the higher eigenmodes  have died out,
$$
c _n (\tau) \approx 0 \hbo \hbox{for} \hbo n \ge i+1 \; ,
$$
while the lower eigenmodes are ``drifting'' along the gradients of
the action: 
$$
c_n(\tau) \; = \; - \, f_n \, \tau \hbo \hbox{for} \hbo n \le i \; .
$$
Hence, by choosing $\tau _\mathrm{stop} \in R$ (\ref{eq:ir11}), it is
possible to separate Gaussian fluctuations from the low-action
collective modes. Those configurations, which are not quite
minima of the actions, are called {\it streamlines} throughout this paper.

\vskip 0.3cm
Although it might be feasible in practice to choose $\tau _\mathrm{stop} \in
R$ hence generating a streamline configuration, the properties of those
streamline  configurations might still depend on the ad hoc
chosen  $\tau _\mathrm{stop}$. It would be highly desirable to replace
the ad hoc stopping of cooling by a physical criterion thereby
linking the properties of the streamlines to a feature of the
theory. {\it Constrained cooling} is precisely delivering that.
To see this, let us return to the flow equation (\ref{eq:ir7})
now with the constraint in place, i.e., $n(x)$ not all zero.
We do not artificially stop cooling and consider the limit
$\tau \to \infty$ (although we admit that this is likely to be
impracticable in an actual computer simulation):
\be
\lim _{\tau \to \infty} \phi (\tau ) \; = \; \phi ^c \; - \; (PMP)^{-1} \,
P \, f \; \not= \; \phi _\mathrm{min} \; .
\label{eq:ir15}
\en
Note that if the constraint is sufficiently weak in the sense
$$
 (PMP)^{-1} \, P \, f \; \approx \; M^{-1} \, f \; ,
$$
the limiting configuration would correspond to a {\it streamline}
configuration of the un-con\-strai\-ned theory.

\vskip 0.3cm
Below, we will detail that the above assumptions are met for the
case of an SU(2) gauge theory where the constraint is in place to
conserve the quark confinement.

\subsection{Constrained cooling in Yang-Mills theory }
\label{sec:ConstrainedCooling}

In Yang-Mills theories, minimal action solutions are the celebrated
instanton configurations which are characterised according to
their integer-valued 
topological charge $Q$. Most relevant at quantum level is
the sector with vanishing topological charge: to this sector,
a gas or liquid of as many instantons ($Q=1$) as anti-instantons ($Q=-1$)
would contribute
besides the empty perturbative vacuum.
Highly relevant properties of the Yang-Mills vacuum
such as the spontaneous breaking of chiral symmetry can be
attributed to the instanton medium. It is well known that  excessive cooling
leads to an annihilation of instanton anti-instanton pairs 
or even to charge changing (anti-)instanton elimination. Since we are 
eventually left with the empty vacuum, cooling has to be stopped to retain 
viable non-perturbative vacuum structure. Here, we will show
that the constrained cooling detailed below results in Yang-Mills
streamline configurations which not only give rise to spontaneous
chiral symmetry, but also to quark confinement. 
The stucture, however, differs from a gas or liquid of instantons.

\vskip 0.3cm
For this purpose, we extract streamline configurations, which are
non-perturbatively relevant, from Monte Carlo configurations occurring
within the lattice formulation. Starting from a snapshot taken from the 
Markoff chain
simulation of the path integral, we propose a constrained minimisation of
the action. We shall describe the procedure for $SU(2)$ Yang-Mills theory
and characterise the emerging field configurations. In the present work
we concentrate on the zero temperature situation (vacuum state) and use a
symmetric $N^4$ Euclidean lattice with periodic boundary conditions and
finite lattice spacing $a$. As usual for $SU(2)$ gauge theory, the gluon
degrees of freedom are represented by unitary matrices
\be
U_\mu (x) \; = \; u_\mu ^0(x) + i \vec{u}_\mu (x) \vec{\tau } , \hbo
u^0_\mu(x)^2 + \vec{u}_\mu(x)^2 \; = \; 1 \; ,
\label{eq:cc1}
\en
associated with the links of the lattice. We are going to work with the
Wilson action
\bea
S_\mathrm{Wilson} &=&  \sum_{x,\, \mu > \nu} \Bigl[ 1 \; - \;
\frac{1}{2} \, \tr \, P_{\mu \nu} (x) \Bigr] \, ,
\label{eq:2a} \\
P_{\mu \nu} (x) &=&
U_\mu (x) \, U_\nu (x+\mu) \, U^\dagger _\mu (x+\nu) \, U^\dagger _\nu (x)
\; .
\label{eq:2b}
\ena
The Yang-Mills partition function is given as the discrete functional integral
\be
Z \; = \; \int {\cal D} U \; \exp \{ - \beta \, S_\mathrm{Wilson} \} \, ,
\label{eq:3}
\en
where $\beta = 4/g^2$ is related to the bare gauge coupling, and represents the
relative importance of gauge field configurations. This fixes also the length
scale by the corresponding lattice spacing $a=f(\beta)$.
The integration measure comprises the Haar measure for each link.
Observables $A(U)$ are calculated by means of a Markoff chain Monte-Carlo
approach which allows to obtain expectation values by importance sampling of
relevant lattice configurations $\{ U_\mu \}$ and averaging over $A(U)$.

\vskip 0.3cm
Starting from such an importance-sampled lattice configuration, a configuration
with reduced action is obtained by a mapping summarising the cooling procedure:
$$
\{ U \}  \stackrel{\mathrm{cooling}}{\longrightarrow } \{ U^c \} \, .
$$
Thereby, we are concentrating on the streamline regime when the Gaussian
fluctuations have been removed already. In practice, we will monitor the
evolution of observables, averaged over an ensemble of Monte Carlo configurations,
as a function of the number $N_{\rm cool}$ of cooling sweeps. It is well known
(and will be demonstrated again below), that if cooling is performed by steepest
descent with respect to the gauge action, confinement fades away with
$N_{\rm cool} \to \infty$.

\vskip 0.3cm
Using the constrained cooling procedure proposed in this paper, our aim is to
generate streamline configurations from the Monte Carlo configurations, such
that an ensemble of them, hopefully, still supports the properties of quark
confinement and chiral symmetry breaking. To this aim, we take a radical
attitude
as announced above and design a cooling procedure which largely reduces the action
while it does not allow to change {\it every single} Polyakov line in all
Euclidean directions on the lattice.
If $P_4$ denotes 
one Polyakov line in time-like direction, $x_4=t$,
$$
P_4(\vec{x}) \; = \; \tr \prod_{t} U_4(t,\vec{x}) , \hbo x = (t,\vec{x})
$$
the static, colour averaged quark-antiquark potential 
$V(r)$ can be obtained via the
correlator of two Polyakov lines
\be
\exp \{ - V(r) N a \} \; = \; \left\langle P_4(\vec{x}) \, P_4(\vec{y})
\right\rangle , \hbo \vert \vec{x} - \vec{y} \vert = r \; ,
\label{eq:4}
\en
where the average $\langle \ldots \rangle $ is over an ensemble of
configurations $\{U \}$. If $\langle \ldots \rangle_c $ denotes the
average over {\it cooled} configurations $\{ U^c \}$, the constraint
trivially implies that the cooled configurations reproduce the static
colour averaged potential
\be
\left\langle P_4(\vec{x}) \, P_4(\vec{y}) \right\rangle_c \; = \;
\left\langle P_4(\vec{x}) \, P_4(\vec{y}) \right\rangle \; = \;
\exp \{ - V(r) N a \} \, ,
\label{eq:5}
\en
that could also be measured on the ensemble of non-cooled configurations.
We will point out that not only the static quark-antiquark potential
is preserved, but also the potential between charges of higher representations.
At zero temperature, the four Euclidean directions should be
equivalent. Therefore we formulate the constraint for cooling such that also
spatial Polyakov loops are not allowed to change. In the result, one can expect
that this constrained cooling reproduces also the area law for fundamental
Wilson loops that gives the colour singlet quark-antiquark potential.
This will be checked below in this subsection.

\vskip 0.3cm
The only way that our original hope could fail is when the constrained cooling
fails to reduce the action by a significant amount. If the cooling sweeps would
be seriously inhibited by the constraints, the action would be reduced only by
an insignificant amount. This would imply that the ``streamline configurations''
we are looking for would not be smooth semiclassical fields.
Whether this is the case or not can be decided by a numerical experiment. 
Below, we will show that the specific constraints do not prevent the cooling 
procedure from largely reducing the action.

\begin{figure}
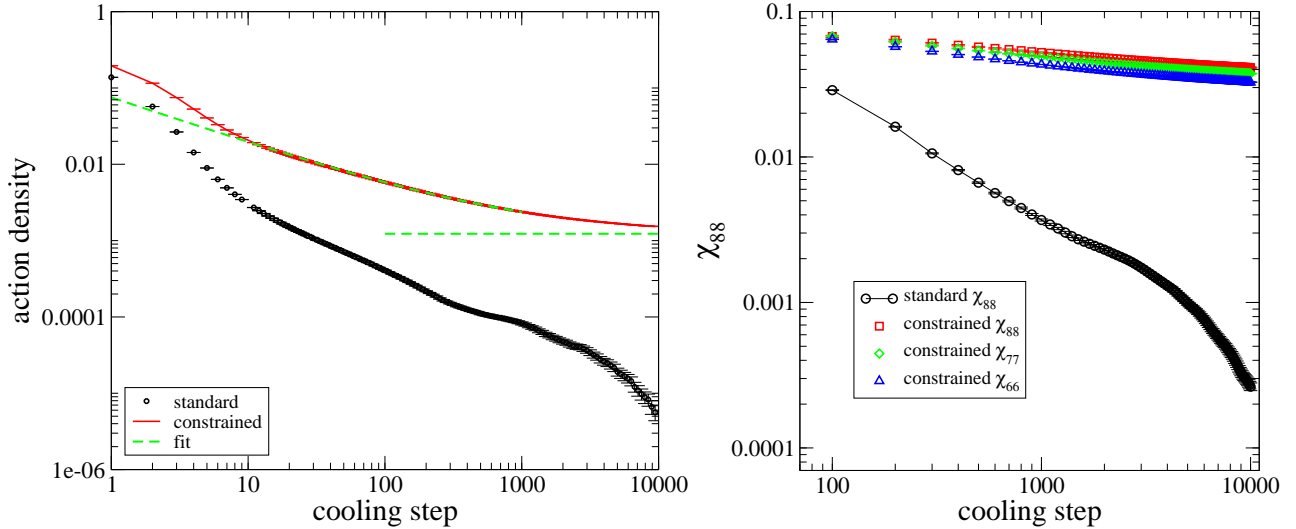

\includegraphics[height=7cm]{action_den3_pap.eps}
\includegraphics[height=7cm]{creutz88_log_pap.eps}
\caption{The action density $\overline{s}$ (left) and some Creutz ratios
$\chi_{nn}$ (right) as a function of the number of (constrained) cooling sweeps 
for an ensemble of $784$ Monte Carlo configurations created at $\beta=2.4$
on a lattice $16^4$. 
For comparison, the case of standard cooling is also shown for $\overline{s}$
and $\chi_{88}$ (1008 configurations). }
\label{fig:1_action_and_Creutz}
\end{figure}
The constrained cooling is easy to formulate. Defining the {\it staples } as usual
by
\be
B_\mu (x) \; = \; \sum _{\nu \not=\mu} \Bigl[ U_\nu (x+\mu) \, U^\dagger _\mu (x+\nu) \,
U^\dagger _\nu (x) + U^\dagger_\nu(x+\mu-\nu) \, U^\dagger_\mu(x-\nu) \,
U_\nu(x-\nu) \, \Bigr] \; ,
\label{eq:6a}
\en
the part $s_\mu (x)$ to which a link $U_\mu (x)$ contributes is given by
\be
s_\mu (x) \; = \; - \, \tr \, \Bigl\{ U_\mu (x) \, B_\mu (x) \Bigr\}  \; .
\label{eq:6}
\en
In order to implement the constraints, we define incomplete Polyakov lines
in any direction $\mu$,
\be
{\cal P}_\mu (x) \; = \; \prod _{n=1}^{N-1} U_\mu(x +n \vec{e}_\mu) \; ,
\label{eq:7}
\en
such that the trace of the Polyakov line through some point $x$ can be
written as
\be
P_\mu(\vec{x}) \; = \; \tr \Bigl\{ U_\mu(x) \, {\cal P}_\mu(x) \Bigr\} , \hbo
\forall \,  x_\mu \; .
\label{eq:8}
\en
In order to preserve the (hypercubic) isotropy we impose the condition
that Polyakov lines remain unchanged to {\it all} directions $\mu$.
The update is done sequentially.
If $U^{(n)}_\mu$ is the link before the update into $U^{(n+1)}_\mu$, and
${\cal P}^{(n)}_\mu(x)$ is the incomplete Polyakov loop with all previous
updates performed, then the following is required to hold:
\be
\tr \Bigl\{ U_\mu(x) \, {\cal P}_\mu (x) \Bigr\} \; = \;
\tr \Bigl\{ U^{(n)}_\mu(x) \, {\cal P}^{(n)}_\mu(x) \Bigr\} \; = \;
\tr \Bigl\{ U^{(n+1)}_\mu(x) \, {\cal P}^{(n)}_\mu(x) \Bigr\} , \hbo
\forall \, x, \, \mu \; , 
\label{eq:10}
\en
where $U^{(0)}_\mu(x) = U_\mu(x) $. 
Thus, the original Polyakov loops are preserved.
The latter condition is built in as a constraint using Lagrange multipliers
$\lambda _\mu(x)$. When a particular link $U_\mu(x)$ becomes subject to
relaxation $U^{(n)}_\mu \to U^{(n+1)}_\mu$, we are led to minimise the
principal action
\be
A_\mu(x) \; = \; s_\mu(x) \, - \, \lambda _\mu(x) \,
\tr \Bigl\{ U_\mu(x) \, {\cal P}^{(n)}_\mu (x) \Bigr\}
\; = \;
- \; \tr \Biggl\{ U_\mu(x) \, \Bigl[ B^{(n)}_\mu(x) + \lambda _\mu (x) \,
{\cal P}^{(n)}_\mu (x) \Bigr] \Biggr\}
\label{eq:11}
\en
with respect to the link $U_\mu(x)$, the result to be stored as part of the
iterated configuration $U^{(n+1)}_\mu$. This is particularly easily done for
the gauge group $SU(2)$. Thereby $B^{(n)}_\mu (x)$ is the staple formed out
of links with all previous updates performed. We find the global minimum of
the principal action for any choice of the Lagrange multiplier to be attained
for
\be
U^{(n+1)}_\mu (x) \; = \; \frac{1}{\cal N} \,
\Bigl[ B^{(n) \, \dagger} _\mu(x) + \lambda _\mu (x) \,
{\cal P}^{(n) \, \dagger} (x) \Bigr] ,
\; \; \;
{\cal N} := \mathrm{det} ^{1/2} \Bigl[ B^{(n)}_\mu(x) + \lambda _\mu (x) \,
{\cal P}^{(n)}_\mu (x) \Bigr] \; .
\label{eq:12}
\en
The remaining task is to fix the Lagrange multiplier such that the constraint
in (\ref{eq:10}) is satisfied. Details of the calculation are left to the
appendix~\ref{sec:appA}. It turns out that there is a unique solution for
$\lambda _\mu$. The final result is
\bea
\lambda _\mu &=&  - \frac{1}{2} \,
\tr (B_\mu^{(n)\, \dagger} {\cal P}^{(n)}_\mu ) \; + \; \frac{1}{2}
\tr (U^{(n)}_\mu {\cal P}^{(n)}_\mu ) \; {\cal N}
\label{eq:15} \\
{\cal N} &=& \sqrt{ \frac{ \mathrm{det} (B^{(n)}_\mu) \, - \, \frac{1}{4}
\tr^2 (B^{(n)\, \dagger}_\mu {\cal P}^{(n)}_\mu) }{ 1 - \frac{1}{4} \tr^2
(U^{(n)}_\mu {\cal P}^{(n)}_\mu) }} \; ,
\label{eq:16}
\ena
or in summary
\be
U^{(n+1)}_\mu (x) \; = \; \frac{1}{\cal N} \,
\Bigl[ B^{(n)\, ^\dagger} _\mu(x) - \frac{1}{2} \, \tr  \Bigl(B^{(n)\, ^\dagger
  }_\mu{\cal P}^{(n)}_\mu  \Bigr) \,  {\cal P}^{(n)\, ^\dagger} _\mu
\, \Bigr] \; + \; \frac{1}{2} \, \tr \Bigl(U^{(n)}_\mu {\cal P}^{(n)}_\mu
\Bigr) \, {\cal P}^{(n)\, ^\dagger} _\mu  \; .
\label{eq:15b}
\en
Here no summation over $\mu $ is implied. This result has an intuitive
interpretation: while the $B^{(n)\, ^\dagger}$ term in the square brackets
minimises the action locally, the remaining terms implement the constraints
by projection.

\vskip 0.3cm
The constrained cooling method works as follows:
\begin{itemize}
\item generate a thermalised configuration of links $\{U_\mu\}=\{U^{(0)}_\mu\}$;
\item randomly choose a link, say $U_\mu (x)$ on the lattice and ``update'' the
link accordingly to local cooling:
$U^{(n)}_\mu (x) \to U^{(n+1)}_\mu (x)$; if this is repeated $4N^4$ times
(equal to the number of links on the lattice !)
this is considered as one cooling sweep;
\item for a sufficiently large (but for practical reasons finite) number
$N_\mathrm{cool}$, we obtain the desired semi-classical configuration.
\end{itemize}
In practice, the quantities calculated with these configurations
will be carefully monitored for a possible residual dependence
on $N_\mathrm{cool}$.

\vskip 0,3cm
We finally mention a technical point: excessive accumulation of rounding errors
can deteriorate the Polyakov line values
in the chain
$$
\tr \Bigl\{ U^{(0)}_\mu(x) \, {\cal P}^{(0)}_\mu(x) \Bigr\} \; \approx \;  \ldots
\; \approx \; \tr \Bigl\{ U^{(n)}_\mu(x) \, {\cal P}^{(n)}_\mu(x) \Bigr\} \;
\not= \; \tr \Bigl\{ U^{(0)}_\mu(x) \, {\cal P}^{(0)}_\mu(x) \Bigr\} 
$$
of equalities.
For $N_\mathrm{cool}$ bigger than 
${\cal O}(10^4)$, we would advise to modify the cooling step (\ref{eq:15b}) by
\bea
U^{(n+1)}_\mu (x) &=& \frac{1}{\cal N} \,
\Bigl[ B^{(n)\, ^\dagger} _\mu(x) - \frac{1}{2} \, \tr  \Bigl(B^{(n)\, ^\dagger
  }_\mu{\cal P}^{(n)}_\mu  \Bigr) \,  {\cal P}^{(n)\, ^\dagger} _\mu
\, \Bigr] \; + \; \frac{1}{2} \, \tr \Bigl(U^{(0)}_\mu {\cal P}^{(0)}_\mu
\Bigr) \, {\cal P}^{(n)\, ^\dagger} _\mu  
\label{eq:15c} \\
{\cal N} &=& \sqrt{ \frac{ \mathrm{det} (B^{(n)}_\mu) \, - \, \frac{1}{4}
\tr^2 (B^{(n)\, ^\dagger} _\mu {\cal P}^{(n)}_\mu) }{ 1 - \frac{1}{4} \tr^2
(U^{(0)} _\mu {\cal P}^{(0)}_\mu) }} .
\nonumber
\ena
It is easy to check that the constraints are exactly satisfied:
$$
\tr \Bigl\{ U^{(n+1)}_\mu(x) \, {\cal P}^{(n)}_\mu(x) \Bigr\} \; = \;
\tr \Bigl( U^{(0)}_\mu(x) \, {\cal P}^{(0)} _\mu(x) \Bigr) \; .
$$

\vskip 0.3cm
Below, we will compare the outcome of our new cooling method with results
obtained with standard (un-constrained) cooling. The standard local cooling
step is given by applying
\be
U^{(n+1)}_\mu (x) \; = \; \frac{1}{\cal N} \,
B^{(n) \dagger } _\mu(x) \, ,
\hbo {\cal N} := \mathrm{det} ^{1/2} B^{(n)}_\mu(x) \; , \hbo
\hbox{(standard)} \;
\label{eq:17}
\en
to a randomly selected link. Again we call $4N^4$ local cooling steps
one cooling sweep. The configuration reached after $N_\mathrm{cool}$
sweeps is called ``cooled configuration''. We label an ensemble of such
configurations by the number of sweeps, $N_\mathrm{cool}$
\be
U^c_\mu(x) \; = \; U^{(N_\mathrm{cool} )}_\mu(x) \; ,
\label{eq:17b}
\en
even though the actual number of updates some particular link has experienced
may differ because of the random selection.

\vskip 0.3cm
We have tested the constrained cooling approach using an equilibrium ensemble
of thermalised configurations generated on a $N^4=16^4$ lattice with Wilson's
one-plaquette action at $\beta = 2.4$. For this inverse coupling the lattice
spacing $a$ in units of the string tension $\sigma $ is roughly given by
$\sigma a^2 = 0.0738(5)$~\cite{Langfeld:2007zw}.
The left panel of Figure~\ref{fig:1_action_and_Creutz} shows the ensemble
average of the space-time averaged action density
\be
\overline{s} = \frac{1}{N^4} \sum_{x} \sum_{\mu < \nu}
\left(1 \, - \, \frac{1}{2} \tr~P _{\mu \nu}(x) \right) \; ,
\en
where $P_{\mu \nu}(x)$ is the plaquette (\ref{eq:2b}),
as a function of the cooling sweep for our constrained cooling method and
for the standard one. It turns out that, despite of the constraint, average
plaquette values very close to $1$ are reached.
We also mention that if the Polyakov lines of a particular configuration 
are not constant over the orthogonal
3-volume, constrained cooling can never settle 
with the empty vacuum since this state is incompatible with varying Polyakov 
lines. Indeed, if we fit the action
density averaged over the ensemble of constrained cooled configurations as a
function of the number $n=N_\mathrm{cool}$ of cooling sweeps, we find empirically
\be
\overline{s} \approx 0.0012(1) + 0.074(1) \; n^{-0.60(1)} \; ,
\hbo \chi^2 = 0.02 \; .
\en
The fit is also shown in figure~\ref{fig:1_action_and_Creutz}
and seems to be a fair representation
of the data for $n \ge 100$.

\vskip 0.3cm
The circumstance that the Polyakov line correlator from the constrained-cooled
configurations equals the correlator from the original Monte Carlo ensemble,
implies that the cooled configurations share the same quark-antiquark 
potential
with the uncooled $SU(2)$ theory. It is also instructive to investigate the
amount of disorder remaining in the Wilson loops. For this purpose, we consider 
the Creutz ratios
\be
\chi _{ts} \; = \;- \; \ln \; \frac{
\Bigl\langle W(t,s)  \Bigr\rangle \, \Bigl\langle W(t-1,s-1) \Bigr\rangle }{
\Bigl\langle W(t-1,s) \Bigr\rangle \, \Bigl\langle W(t,s-1) \Bigr\rangle }
\; ,
\label{eq:18}
\en
where $W(t,s)$ are rectangular Wilson loops with extension $t$ in time
and extension $s$ in spatial direction. We consider here only diagonal Creutz
ratios $\chi _{nn}$. For $n$ such large that the excited states have died out, 
the Wilson loop expectation factorises into an exponential
term with circumference and area term and a $t$-independent 
overlap matrix element.
In this case,  the Creutz ratios exhibit only the area term and approach the
asymptotic force from below as a function of $n$
\be
\lim _{n \to \infty} \chi _{nn} \; = \; \sigma \, a^2 \; ,
\en
where  $\sigma a^2$ is the string tension in units of the lattice spacing.
The right panel of Figure~\ref{fig:1_action_and_Creutz} shows some Creutz
ratios $\chi_{nn}$ as functions of the cooling sweep. We observe that 
$\chi_{nn}$ approaches the string tension for an increasing number $n$, 
as it must be. In fact, we observe for constrained cooling
that the string tension is approached from below with increasing $n$.
Without any cooling the Creutz ratios approach the string tension
from above. The reason for this apparent discrepance is easily understood 
by noting that the Creutz ratio can be interpreted as the derivative of the
static quark potential,
$$
\chi _{nn} \approx a^2 \, \frac{dV}{dr} \Big\vert _{r=r_n} \approx
\frac{1}{n^2} \, + \, \sigma a^2 .
$$
Since cooling efficiently wipes out the Coulombic part of the quark potential, 
there is no reason anymore that Creutz ratios tend to overestimate the string 
tension. 
For $\chi _{nn}$ at $n=8$, we still observe a slight deviation from the 
un-cooled value. We attribute this observation to a decrease of the gap
between  between ground state and excited state energies during cooling. 
In this case, even larger values of $n$ (and, in fact, larger lattices) 
would be neccessary to recover the asymptotic value. 

\vskip 0.3cm
Standard cooling yields a quite different picture.
There $\chi _{88}$ is rapidly decreasing and the asymptotic value is
compatible with zero.
This confirms the common wisdom that configurations which arise from standard
cooling do not sustain quark confinement.

\subsection{How good is the semi-classical expansion?}
\label{sec:GluonicSpectra}

\begin{figure}
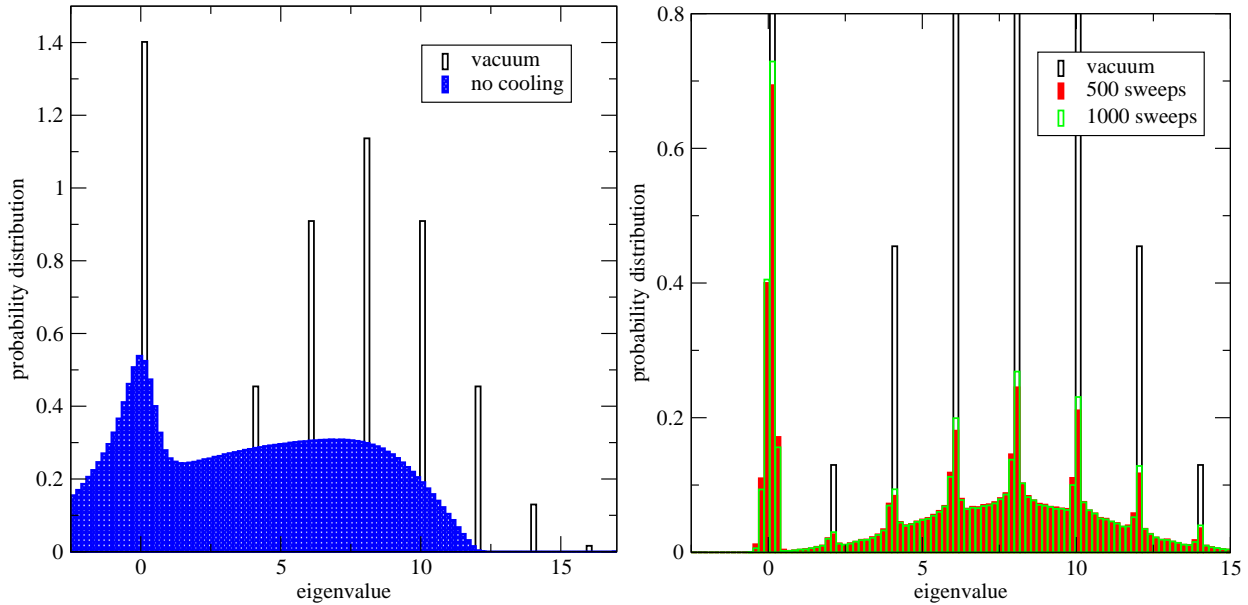

\includegraphics[height=8cm]{prob4x4_bare_pap.eps}   
\includegraphics[height=8cm]{prob4x4_cool5_pap.eps}  
\caption{Left: the spectral density of the Hessian as a function
of its eigenvalues for the perturbative vacuum and the average spectrum
from un-cooled configurations. Right: the spectral density {\it after}
500 and 1000 sweeps of constrained cooling. Results are shown for 320
configurations of a $4^4$ lattice at $\beta=2.3$ and one configuration
corresponding to $\beta=\infty$ (perturbative vacuum).}
\label{fig:2_spectra_4x4x4x4}
\end{figure}
Since we have discovered above low action constrained solutions which
support important features of Yang-Mills theory such as quark confinement,
the question arises whether a semi-classical expansion around these
configurations could be a reliable tool
to describe the theory even quantitatively.
As detailed by Huang in~\cite{Huang:1996wj} and explained in
subsection~\ref{sec:Streamlines}, a first indication would be
finding a gap in the eigenvalue spectrum of the Hessian.
In this subsection, we will show that a clear separation between
Gaussian fluctuations (in the bulk) and low lying collective modes 
is clearly observed for our streamline background configurations.

\vskip 0.3cm
In the context of gauge theories, the parameterisation of the fluctuations
must be carefully chosen 
in order to obtain eigenvalues of the Hessian which are invariant with respect
to gauge transformations of the background field. A suitable parameterisation
was given by Luescher and Weisz~\cite{Luscher:1995vs} and will be adopted here:
if $\{U^c_\mu \}$ denotes the streamline configuration, i.e., the background
field, in the neighbourhood a particular link is decomposed into
\be
U_\mu (x) \; = \; \exp \{ i \, q_\mu (x) \, \} \,
U^c_\mu (x) \; , \hbo q_\mu (x) = q^a_\mu (x) \tau ^a  \; ,
\label{eq:sa1}
\en
where the $q_\mu^a(x)$ represent the fluctuations. These fields change under a
gauge transformation $\Omega(x)$ as follows:
\bea
U^\Omega_\mu (x) &=& \Omega (x) \; U_\mu (x) \; \Omega ^\dagger (x+\mu) \, ,
\nonumber \\
U^{c \; \Omega } _\mu (x) &=& \Omega (x) \; U^c_\mu (x) \; \Omega ^\dagger (x+\mu) \, , \\ 
q_\mu^\Omega (x) &=& \Omega (x) \; q_\mu (x) \; \Omega ^\dagger (x) \; .
\label{eq:sa3}
\ena
Gauge transformations simply rotate the fluctuation field $q^a_\mu(x)$:
\bea
q^{a \; \Omega}_\mu (x) &=&  R^{ab}(x) \, q^{b}_\mu (x) \; , \nonumber \\
R^{ab}(x) \, \tau ^b &=& \Omega (x) \, \tau ^a \,\Omega ^\dagger (x) \; .
\label{eq:sa4}
\ena
\begin{figure}
\includegraphics[height=8cm]{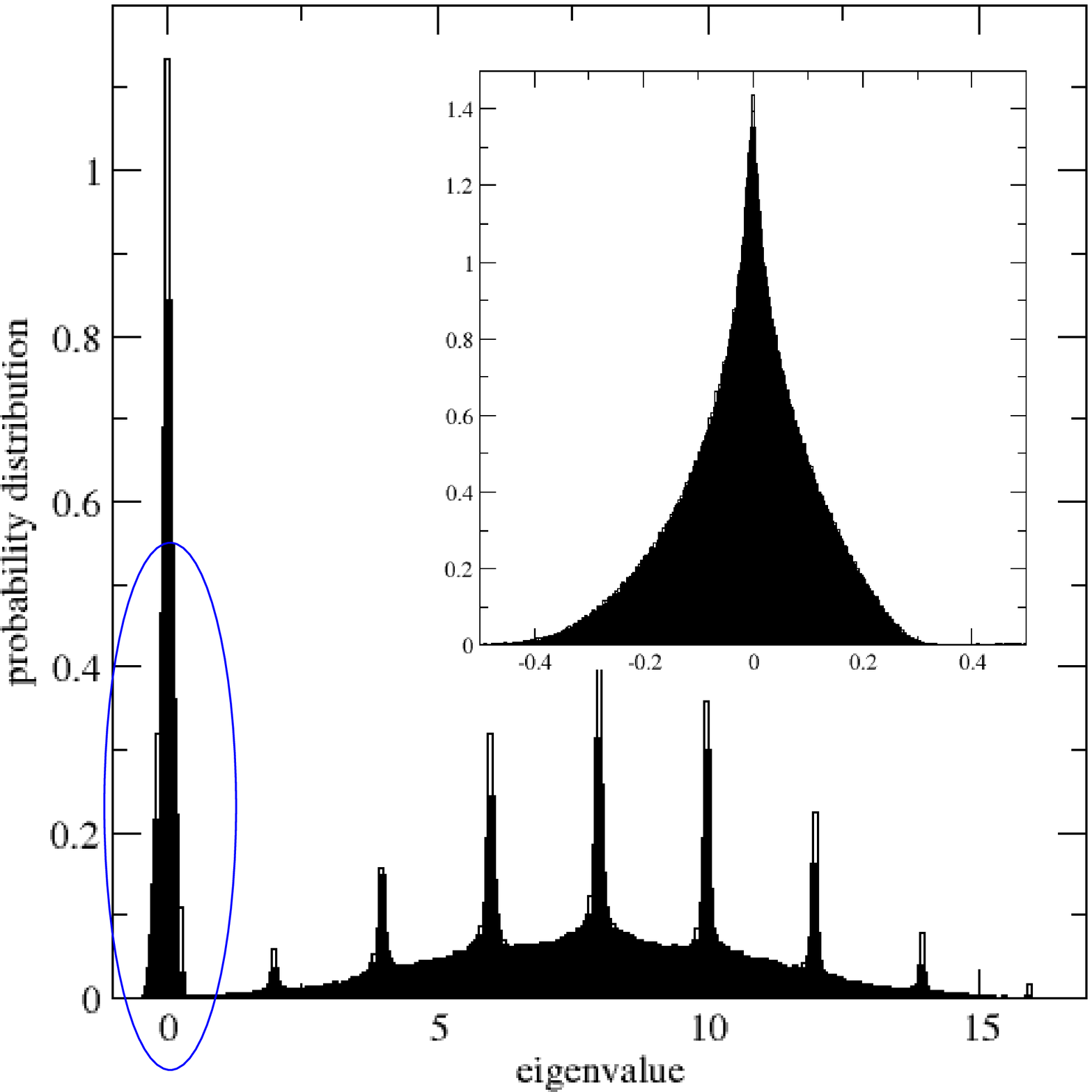}
\includegraphics[height=8cm]{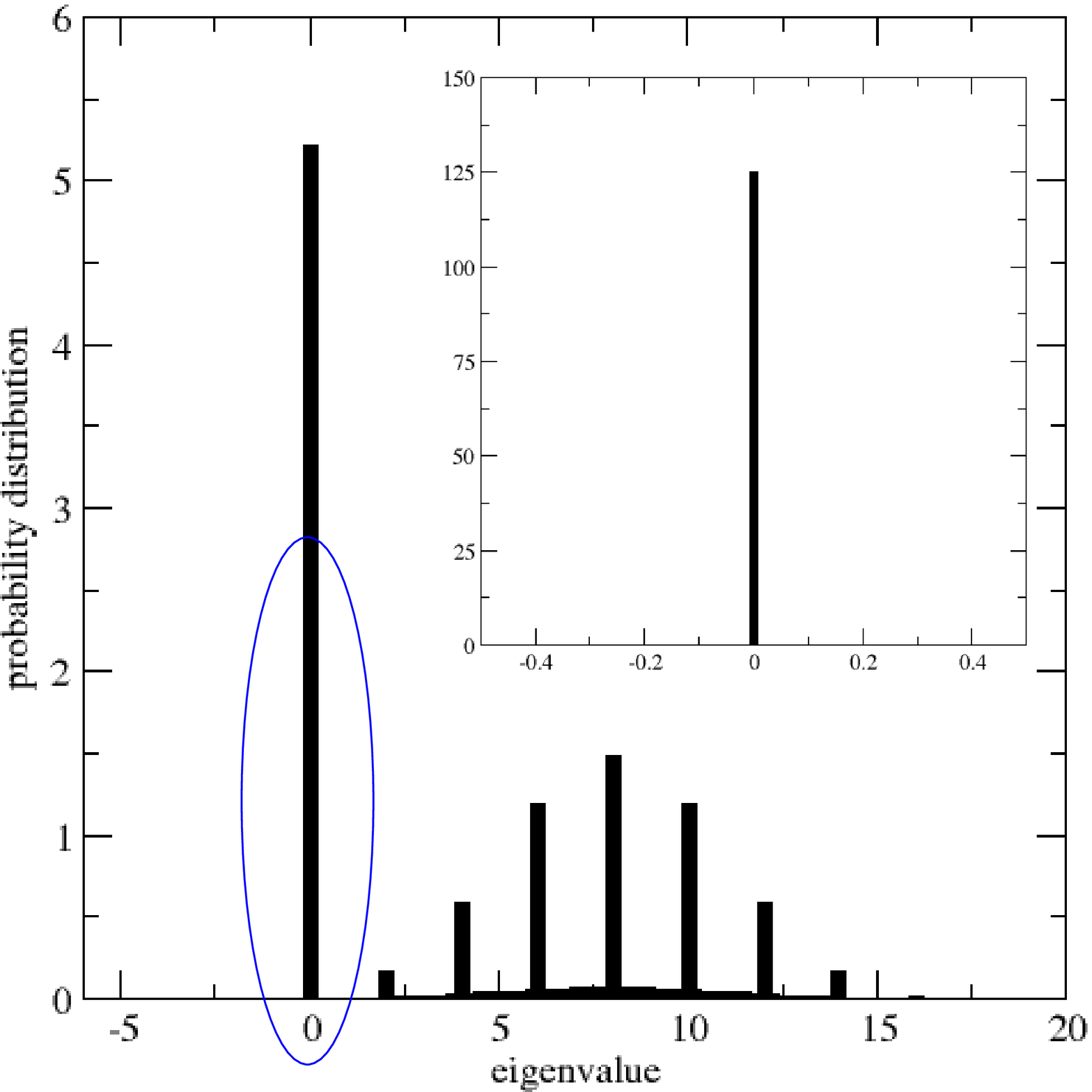}
\caption{Left: the spectral density of the Hessian as a function
of its eigenvalues after 10000 sweeps of constrained cooling.
Right: the spectral density after the same number of standard cooling
sweeps. The densities are the result of averaging over 320 independent
$4^4$ configurations.}
\label{fig:3_constrained_vs_standard_4x4x4x4}
\end{figure}
Expanding the Yang-Mills action up to second order in the fluctuations
yields:
\bea
S[U] &=& S[U^c] \; + \; \sum_{x,a \mu} q^a_\mu (x) \, f^a_\mu[U^c]
\; + \; \frac{1}{2} \sum _{xy,ab,\mu \nu} q^a_\mu (x) \; M^{ab}_{\mu \nu} (x,y)
\; q^b_\nu (y) \; + \; {\cal O}(q^3) \; ,
\label{eq:sa5}
\ena
where $f^a_\mu$ is the drift force and $M^{ab}_{\mu \nu} $
is the Hessian. Because the action is gauge invariant and in view of
the transformation properties (\ref{eq:sa3}),
one easily finds that the Hessian transforms homogeneously under
a gauge transformation of the background field:
\be
M^{ab}_{\mu \nu} (x,y) [U^{c \; \Omega} ] \; = \;
R^{ac}(x)  \; M^{cd}_{\mu \nu} (x,y) [U^c] \; (R^T)^{db}(y) \; .
\label{eq:sa6}
\en
This implies that the eigenvalues of the Hessian are gauge invariant,
and the corresponding eigenmodes are a mere rotation of the
eigenvectors of $M^{ab}_{\mu\nu}[U^c]$.

\vskip 0.3cm
For the Wilson action (\ref{eq:2a}) and with the definition
$$
\delta ^a U_\mu (x) \; = \; i \, \tau^a \, U_\mu (x) \; , \hbo
\delta ^a U^\dagger_\mu (x) \; = \; [ \delta ^a U_\mu (x) ]^\dagger \; = \;
-i \, U^\dagger_\mu (x) \, \tau^a \; ,
$$
the Hessian is given by:
\bea
M^{ab}_{\mu \nu} (x,y) &=&
\frac{1}{2} \, \tr \Bigl\{ U_\mu (x) B_\mu (x) \Bigr\} \;
\delta ^{a,b} \; \delta _{x,y} \; \delta _{\mu,\nu }
\nonumber \\
&-& \frac{1}{2} \, \Bigl\{ \delta ^a U_\mu(x) \, \delta ^b U_\nu(x+\mu) \,
U^\dagger _\mu (x+\nu) \, U^\dagger _\nu (x) \Bigr\} \;
\delta _{y,x+\mu } \; [ 1 - \delta _{\mu,\nu} ]
\nonumber \\
&-& \frac{1}{2} \, \Bigl\{ \delta ^a U_\mu(x) \, U_\nu(x+\mu) \, \delta ^b
U^\dagger _\mu (x+\nu) \, U^\dagger _\nu (x) \Bigr\} \;
\delta _{y,x+\nu } \; \delta _{\mu,\nu}
\nonumber \\
&-& \frac{1}{2} \, \Bigl\{ \delta ^a U_\mu(x) \, U_\nu(x+\mu) \,
U^\dagger _\mu (x+\nu) \, \delta ^b U^\dagger _\nu (x) \Bigr\} \;
\delta _{y,x} \; [ 1 - \delta _{\mu,\nu} ]
\nonumber \\
&-& \frac{1}{2} \, \Bigl\{ \delta ^a U_\mu(x) \, \delta ^b
U^\dagger_\nu(x+\mu-\nu ) \, U^\dagger _\mu (x-\nu) \, U_\nu (x-\nu)
\Bigr\} \;  \delta _{y,x+\mu-\nu} \; [ 1 - \delta _{\mu,\nu} ]
\nonumber \\
&-& \frac{1}{2} \, \Bigl\{ \delta ^a U_\mu(x) \, U^\dagger_\nu(x+\mu-\nu ) \,
\delta ^b U^\dagger _\mu (x-\nu) \, U_\nu (x-\nu)
\Bigr\} \;  \delta _{y,x-\nu} \; \delta _{\mu,\nu} 
\nonumber \\
&-& \frac{1}{2} \, \Bigl\{ \delta ^a U_\mu(x) \, U^\dagger_\nu(x+\mu-\nu ) \,
U^\dagger _\mu (x-\nu) \, \delta ^bU_\nu (x-\nu)
\Bigr\} \;  \delta _{y,x-\nu} \; [ 1 - \delta _{\mu,\nu} ] \; .
\nonumber \\
\ena
The spectral density $\rho(\lambda)$ is defined by the number
$\rho(\lambda) \, d \lambda $ of eigenvalues $\lambda$ of the Hessian in the
interval between $\lambda$ and $\lambda + d\lambda$. We have investigated this
spectral density over the whole range of eigenvalues for a small lattice of
size $4^4$ since the gross features of the spectrum can be already illustrated
using rather small volumes.

\vskip 0.3cm
Let us firstly study the empty, i.e. perturbative, vacuum where
all links are set to unity, $U_\mu (x) =1 $, $\forall x , \, \mu $.
The result is shown by the open bars in the left panel of
Figure~\ref{fig:2_spectra_4x4x4x4}.
The spectrum resembles that of the Laplacian operator {\it besides} of a rather
large number of zero modes. This shows that the empty vacuum possesses a huge
number of {\it flat directions}. This has been previously pointed out
in~\cite{Keurentjes:1998uu,Schaden:2004ah,Langfeld:2005mz}.
In order to imagine one of these flat directions, let us assume that all 
time-like links pointing forward from a given time slice is set to an arbitrary
$SU(2)$ element, let us say $U$. In this case, all plaquettes on the lattice
are still equal to $1$. In other words, this configuration is degenerate
with what we have called perturbative vacuum. On the other hand, 
the Polyakov line of the new configuration is given by $\tr~U$, which is
different from $\tr~1$ of the perturbative vacuum. Since the Polyakov line
is gauge invariant, the new configuration is {\it not} gauge equivalent to our
starting configuration. Variation of the action with respect to the degrees
of freedom of $U$ hence will produce genuine zero modes of the Hessian.

\vskip 0.3cm
In the left panel of Figure~\ref{fig:2_spectra_4x4x4x4},
the full bars show the spectral density
averaged over 320 independent configurations from a thermalised $4^4$ lattice
obtained at $\beta = 2.3$. The discrete spectrum from the perturbative vacuum
is now washed out. In addition, negative eigenvalues naturally appear since a
thermalised configurations does not represent a local minimum of the action.

\vskip 0.3cm
The right panel of Figure~\ref{fig:2_spectra_4x4x4x4} show the spectral
densities corresponding
to different numbers of {\it constrained cooling} sweeps. Note that there is very
little change if the number of cooling sweeps is increased from 500 to 1000.
As expected, the modes which are present for a empty vacuum background get
emphasised during cooling. More importantly, there is a certain amount of near
zero, i.e. collective modes which seem to remain stable under constrained cooling.
Among those there are also negative eigenvalues of the Hessian. Those modes would
normally exponentially grow, destroying the streamline configuration. It is this
growth which is inhibited by the constraints 
imposed upon the cooling process.

\vskip 0.3cm
Note that there is also a gap between the collective low-lying modes and the
higher modes. The latter we therefore identify with the fluctuations, while
the collective modes represent collective deformations of the streamline
configuration. It is interesting to compare the spectrum obtained in the result
of {\it constrained cooling} with that calculated for configurations emerging
under {\it standard cooling}. Our spectral results after $10000$
cooling sweeps in either case are compared in the left and right panels of
Figure~\ref{fig:3_constrained_vs_standard_4x4x4x4}.
The inlays show the eigenvalue spectrum near the peak at
zero in finer resolution (implying a change in the scale due to the fixed
normalisation of the spectral distribution). In sharp contrast to constrained
cooling, the collective modes disappear after a sufficiently large amount of
standard cooling.

\vskip 0.3cm
\begin{figure}
\centering
\includegraphics[height=8cm]{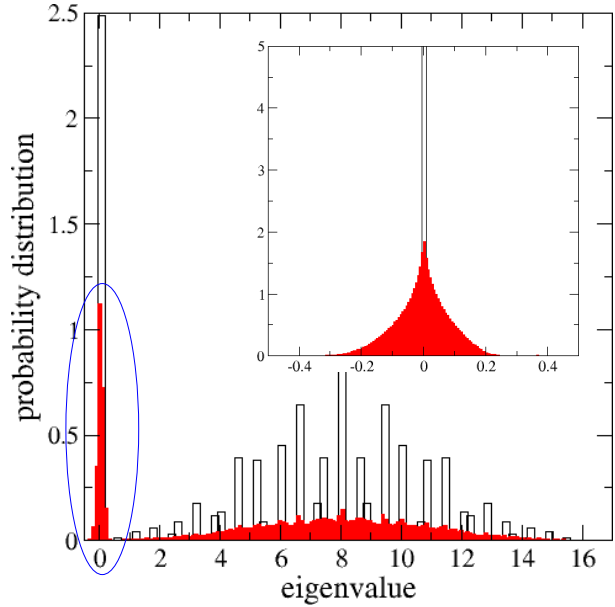}
\caption{\label{fig:h4} The spectral density of the Hessian as a function
of its eigenvalues for a single configuration prepared at $\beta =2.3$
on a $8^4$ lattice, after $1000$ sweeps of constrained cooling. The insert
shows the vicinity of zero.}
\label{fig:4_constrained_for_8x8x8x8}
\end{figure}
The gap between the collective near-zero modes and the Gaussian fluctuations
at intermediate eigenvalues in the bulk spectrum is a generic feature of
constrained cooling. 
According to the inspiring paper by Huang~\cite{Huang:1996wj}, both
the presence of near-zero modes and the spectral gap between them and the bulk
spectrum are the hallmark of semiclassical configurations.
Admittedly, the $SU(2)$ gauge theory with $\beta=2.3$ using an $4^4$ lattice
might be close to the finite-volume pseudo-deconfinement transition.
We therefore checked the spectrum of the Hessian for the same $\beta$ value
using an $8^4$ lattice. This setting comfortably belongs to the confinement
phase of the theory. In order to obtain {\it all} eigenvalues of the Hessian,
the matrix has been exactly diagonalised using the $256$ nodes of the dedicated
cluster at the HPCC, Plymouth. Already the eigenvalue spectrum from a single
lattice configuration shows the familiar pattern (see figure~\ref{fig:h4}):
a clear separation between collective and Gaussian modes.

\section{Space-time texture }
\label{sec:SpaceTimeTexture}

\subsection{Action and topological charge}

Since standard cooling and its refinements (improved action, restricted
cooling) have been developed with the original 
intention to analyse deeper the vacuum
texture, an inhomogeneous structure visible in action and topological charge,
it is clear that we also have to further analyse the space-time structure which
characterises the configuration after constrained cooling. A particularly 
important questions is whether the emerging streamline configurations are 
related in any way to instantons, the apparent structures believed to be seen 
in some stage of standard cooling.

\vskip 0.3cm
\begin{figure}
\includegraphics[height=7cm]{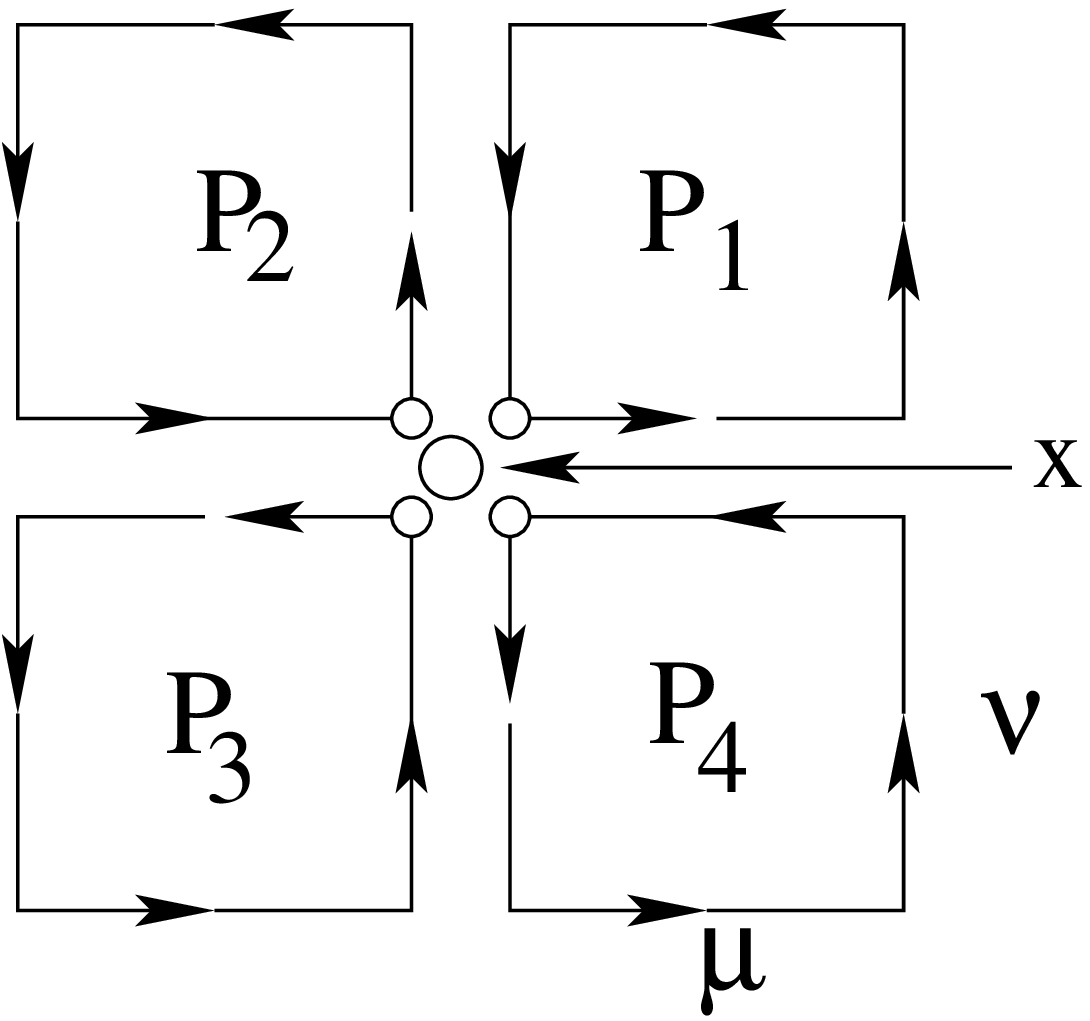} \hspace{0.5cm}
\includegraphics[height=8cm]{topo_dens2_pap.eps}
\caption{Left: the field strength defined as the clover average over plaquettes
touching in the point $x$: $C_{\mu \nu }(x) = (P_1 + P_2 + P_3 + P_4 )/4$.
Right: the topological susceptibility $\chi_{\rm top}$ as a function of the
cooling sweep for constrained and standard cooling.
The configurations have been prepared on a $16^4$ lattice at $\beta = 2.4$. }
\label{fig:5_clover_and_top_susc}
\end{figure}
To define the notation, we briefly review the basic properties of an
1-instanton configuration. In the continuum formulation, the gluon field
and the field strength tensor are given by (in the regular gauge)
\bea
A^a_{\mu}(x) &=& \eta ^a_{\mu\nu}  \, x^\nu \; \frac{ 2}{x^2+\rho^2} \; , \\
F^a_{\mu\nu}(x) &=&  - \; \eta ^a_{\mu\nu}  \; \frac{ 4 \rho ^2 }{[x^2+\rho^2]^2}
\; , 
\label{eq:in1} 
\ena
with $\eta ^a_{0i} = - \eta ^a_{i0} = \delta _{ai}$ and
$\eta ^a_{ik} = \epsilon _{aik}$ . 
The action of the instanton is independent of the instanton radius
$\rho$ and is given by
\be
S_\mathrm{1-instant} \; = \; \frac{1}{4g^2} \sum_{\mu \nu a} \int d^4x \;
F^a_{\mu \nu}(x) F^a_{\mu \nu}(x) \; = \; \frac{8\pi^2}{g^2} \; .
\label{eq:in3}
\en
The (anti-)instanton is an (anti-)selfdual solution of the Yang-Mills equation
of motion:
\be
F^a_{\mu \nu}(x) \, = \, \pm \, \widetilde{F}^a_{\mu \nu}(x) ,
\hbo  \widetilde{F}^a_{\mu \nu}(x) = \frac{1}{2} \epsilon _{\mu \nu
\alpha \beta } \, F^a_{\alpha \beta}(x) \; .
\label{eq:in4}
\en
We also introduce for convenience the colour electric and colour magnetic
fields $(i,k,l \in \{1,2,3\})$:
\be
F^a_{0i} (x) \; = \; E^a_i (x) \; , \hbo F^a_{ik} (x) \; = \; \epsilon _{ikl}
\, B^a_l (x) \; .
\label{eq:in4b}
\en
Smooth configurations composed out of (anti-)instantons are characterised by
their topological charge, expressible in terms of the field strength by
\be
Q = \frac{1}{64\pi^2} \sum_{\mu,\nu,\alpha,\beta} \sum_a \int d^4x 
\epsilon _{\mu\nu\alpha\beta } \, F^a_{\mu\nu}(x)  F^a_{\alpha\beta }(x)
 = \int d^4x \; q(x) \; ,
\en
with the topological density
\be
q(x) \; = \; \frac{1}{8\pi^2} \sum_{i=1,3;a} \;  E^a_i(x) B^a_i(x) \; . 
\label{eq:in5}
\en
We have used the standard notation
$F_{\mu\nu } := F^a_{\mu\nu} \, t^a$ and 
$\tr ( t^a t^b) = \frac{1}{2} \delta^{ab}$ ,
where $t^a$ ($ = \lambda_a/2$ for $SU(3)$) are the generators of the $su(N_c)$ 
Lie algebra.
The action and the topological charge of a dilute gas of $n_I$ instantons
and $\bar{n}_I$ anti-instantons are given by
\be
S \; = \; \frac{8\pi^2}{g^2} \, \Bigl( n_I +  \bar{n}_I \Bigr), \hbo
Q \; = \;  n_I -  \bar{n}_I \; .
\label{eq:in6}
\en
For convenience, we also introduce the ``normalised'' action $A$ by
\be
A \; = \; \frac{1}{16\pi^2}  \sum_{i \, a} \int d^4x \;  \Bigl(
E^a_i (x) E^a_i (x) \, + \, B^a_i (x) B^a_i (x) \; \Bigr) \; ,
\label{eq:in6b}
\en
with the density
\be
s(x) = \frac{1}{16\pi^2} \Bigl( E^a_i (x) E^a_i (x) \, + \, B^a_i (x) B^a_i (x) \; \Bigr)
\en
which gives rise to the global inequality
\be
A \; \ge \; \vert Q  \vert 
\label{eq:inequality}
\en
and a local variant of it:
\be
- s(x) \le q(x) \le s(x)  \; . 
\label{eq:bound}
\en

\begin{table}
\begin{minipage}{0.5\textwidth}
\scriptsize
\begin{tabular}{c|ccc}
 lump number & $N_\mathrm{lump}$ & $Q_\mathrm{lump}$ & $A_\mathrm{lump}$  \\ \hline
   1 &       4251  &  0.6318327E+00 &   0.3244866E+01 \\
   2 &       5123  &  0.1609389E+00 &   0.1796208E+01 \\
   3 &        509  &  0.5354339E+00 &   0.1027845E+01 \\
   4 &         82  &  0.3428497E+00 &   0.3697114E+00 \\
   5 &          1  &  0.1496636E-03 &   0.4925571E-03 \\
   6 &          1  & -0.1994387E-03 &   0.4756932E-03 \\
   7 &          1  &  0.3741456E-03 &   0.4675257E-03 \\
   8 &          1  & -0.7509839E-04 &   0.4525436E-03 \\
   9 &          1  & -0.3260709E-03 &   0.4070134E-03 \\
  10 &          3  &  0.1876750E-03 &   0.3783614E-03
\end{tabular}
\end{minipage}
\hspace{0.5cm}
\begin{minipage}[r]{0.5\textwidth}
\scriptsize
\begin{tabular}{c|ccc}
 lump number & $N_\mathrm{lump}$ & $Q_\mathrm{lump}$ & $A_\mathrm{lump}$  \\ \hline
   1  &      1600  & -0.1349146E+01  &  0.1967493E+01  \\
   2  &      1536  &  0.7005824E+00  &  0.1241354E+01 \\
   3  &       130  & -0.6418117E+00  &  0.6956997E+00 \\
   4  &      2841  & -0.4728359E+00  &  0.1499717E+01 \\
   5  &      2077  &  0.2547383E+00  &  0.9045169E+00 \\
   6  &        17  &  0.1862375E-02  &  0.8850131E-02 \\
   7  &        12  &  0.1632629E-02  &  0.9739530E-02 \\
   8  &        44  &  0.1126855E-02  &  0.1116815E-01 \\
   9  &        10  &  0.1124477E-02  &  0.7585276E-02 \\
  10  &        50  & -0.1077576E-02  &  0.1420463E-01
\end{tabular}
\end{minipage}
\caption{Typical lump structures on a $10^4$ lattice, originally prepared 
at $\beta =2.3$, after 1000 constrained cooling sweeps.
Left: action-based definition with a threshold $t=20\%$ .}
Right: charge-based definition with a threshold $t=10\%$ .
\label{tab:1}
\end{table}

\begin{figure}
\includegraphics[height=5.5cm]{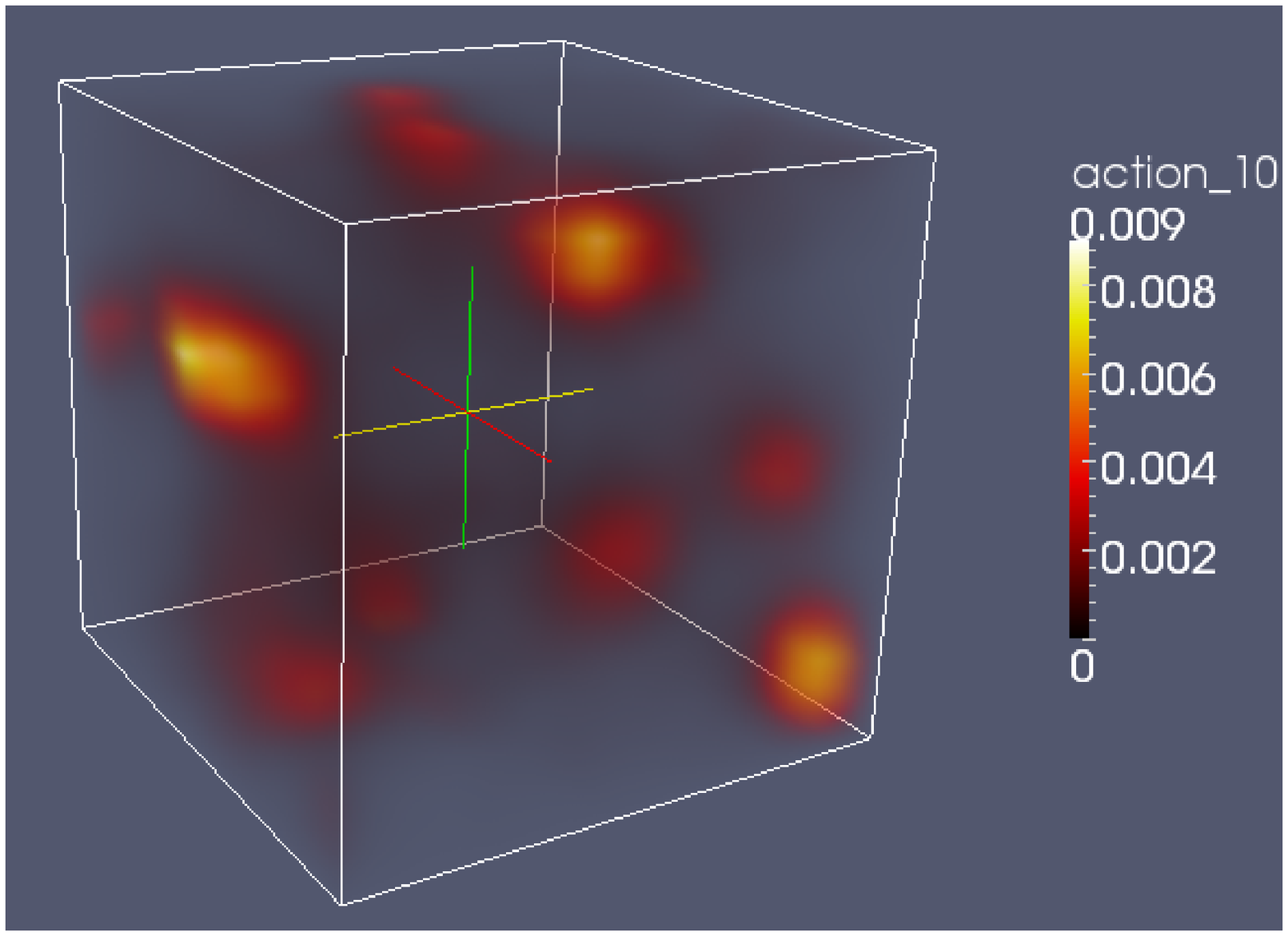} \hspace{0.5cm}
\includegraphics[height=5.5cm]{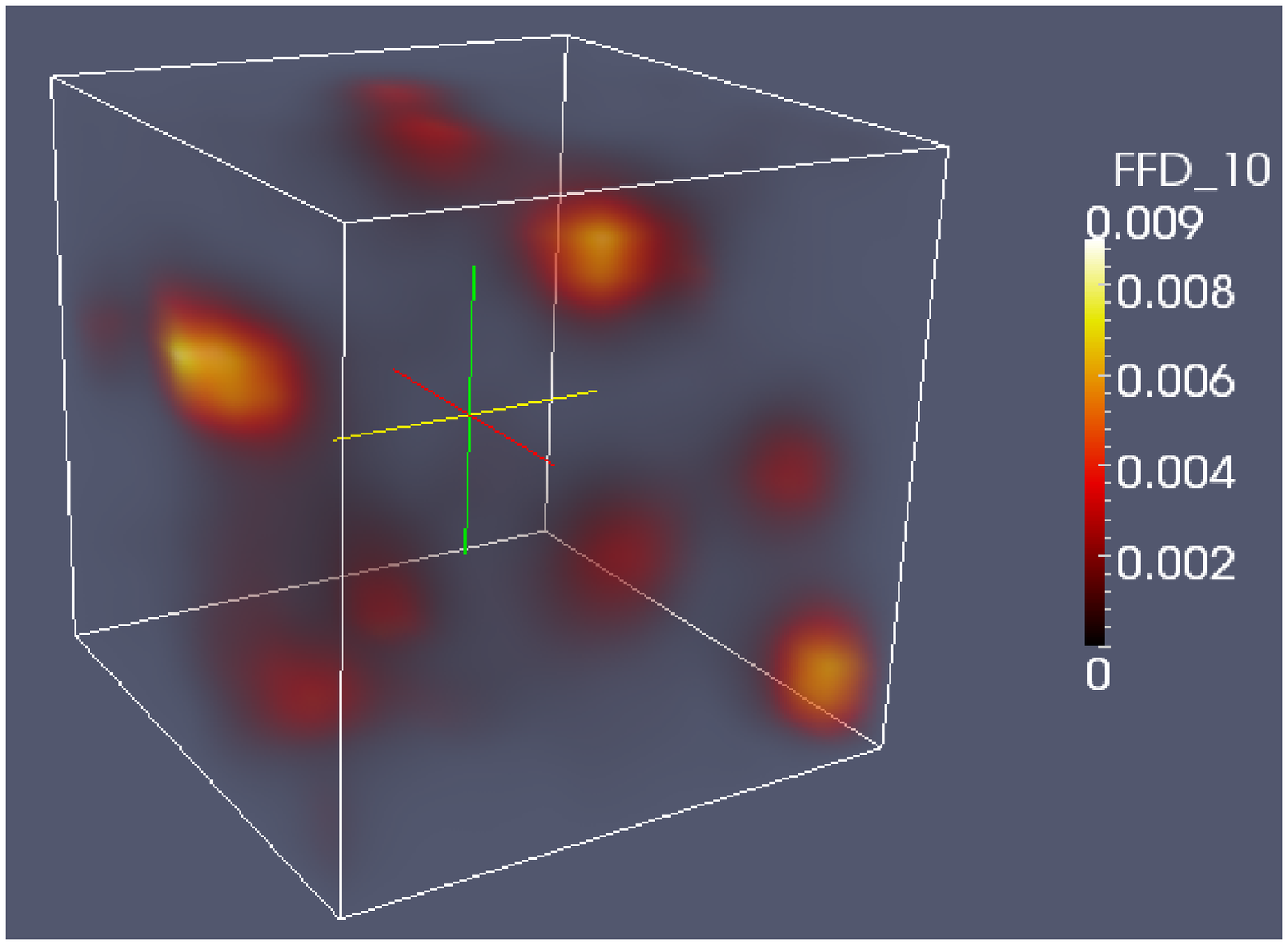}
\caption{Sample configuration obtained after $1000$ constrained cooling
sweeps applied to a Monte Carlo configuration originally prepared on an 
$16^4$ lattice at $\beta = 2.4$.
Left: action density. Right: modulus of the topological charge density.}
\label{fig:6_action_and_topcharge_visual}
\end{figure}
Let us briefly write down the lattice analog of the above quantities.
Since this paper presents a first exploration of a new cooling procedure, we
decided to use the most unambitious (unimproved) lattice observables.
This is not doomed to failure since our cooled configurations are quite
smooth. For example, there is no need for an improved definition of the
field strength which uses other Wilson loops rather than the plaquette.
In order to associate the field strength tensor with each lattice site,
we introduce the clover-leaf average of the plaquettes attached to this
site as indicated in the left panel of Figure~\ref{fig:5_clover_and_top_susc}.
Each of the quantities $P_{1 \ldots 4}$ represents an untraced plaquette 
as defined in (\ref{eq:2b}) with the difference that tadpole improved links
$$
U_\mu (x) / u_0 \; , \hbo u_0 \; = \; \left( \left\langle \frac{1}{N_c} \, \tr
\, P_{\mu \nu } \, \right\rangle \right) ^{1/4}
$$
were used in place of the standard links. For smooth configurations,
i.e. if $a^2 \, F_{\mu\nu }(x) \ll 1$, the clover average $C_{\mu\nu}(x)$
can be expanded with respect to the field strength:
\be
C_{\mu\nu }(x) \; = \; 1 \; + \; i\, a^2 \, F^a_{\mu\nu}(x) \, t^a \; + \;
\ldots \; . 
\label{eq:in7}
\en
It is the latter equation which we use to define the colour components of the
field strength tensor on the lattice:
\be
a^2 \, F^a_{\mu\nu}(x) \; = \; -\, 2\, i \; \tr \Bigl\{ t^a \,
C_{\mu\nu}(x) \Bigr\} \; .
\label{eq:in10}
\en
As a first check, we investigated whether our cooled configurations
still sustain the $U_A(1)$ anomaly. To this aim, we calculated
the topological susceptibility
\be
\chi \; = \; \frac{1}{V} \; \Bigl\langle Q^2 \Bigr\rangle
\label{eq:in11}
\en
where the average is over configurations obtained after constrained
cooling from the Monte Carlo ensemble. The result for 400 independent
lattice configurations of size $16^4$ originating from a simulation at
$\beta = 2.4$ is shown in the right panel of Figure~\ref{fig:5_clover_and_top_susc}
as a function of the cooling
iteration number up to 1000 sweeps. We observe that the maximal value is
reached after ${\cal O}(50)$ cooling sweeps. The asymptotic value is not
essentially less. Using the string tension $\sigma = [440 \, \mathrm{MeV}]^2$
to set the reference scale, our estimate for the asymptotic value
(derived from the largest number of cooling steps) is
\be
\chi ^{1/4} \; = \; (168 \, \pm \, 3) \; \hbox{MeV} 
\label{eq:in12}
\en
is somewhat larger than the value reported in one of
the earliest investigations~\cite{Teper:1985ek} for the $SU(2)$ gauge theory.
There, minimising the action by a variant of the Metropolis algorithm 
(formally taking $\beta \to \infty$) until the (first) plateau value of the 
action is reached, together with the twisted-plaquette definition of the topological 
charge density have allowed to estimate the topological charge of lattice 
configurations. 
Given the lattice sizes of the early simulations, we deem 
our result to be in agreement with the previous results for a $SU(2)$ gauge theory.

\vskip 0.3cm
Figure~\ref{fig:6_action_and_topcharge_visual} visualises the action
density (left panel) and the modulus
of the topological charge density (right panel) of a generic $16^4$ lattice
configuration after $1000$ sweeps of constrained cooling.
We stress two important features: (i) both densities are well localised
in the same, in space-time well separated objects;
(ii) at first sight, the clusters seem to be (anti-) selfdual.
We will quantify the selfduality, more precisely the degree of saturation
of (\ref{eq:bound}), below.

\begin{figure}
\includegraphics[width=0.51\linewidth]{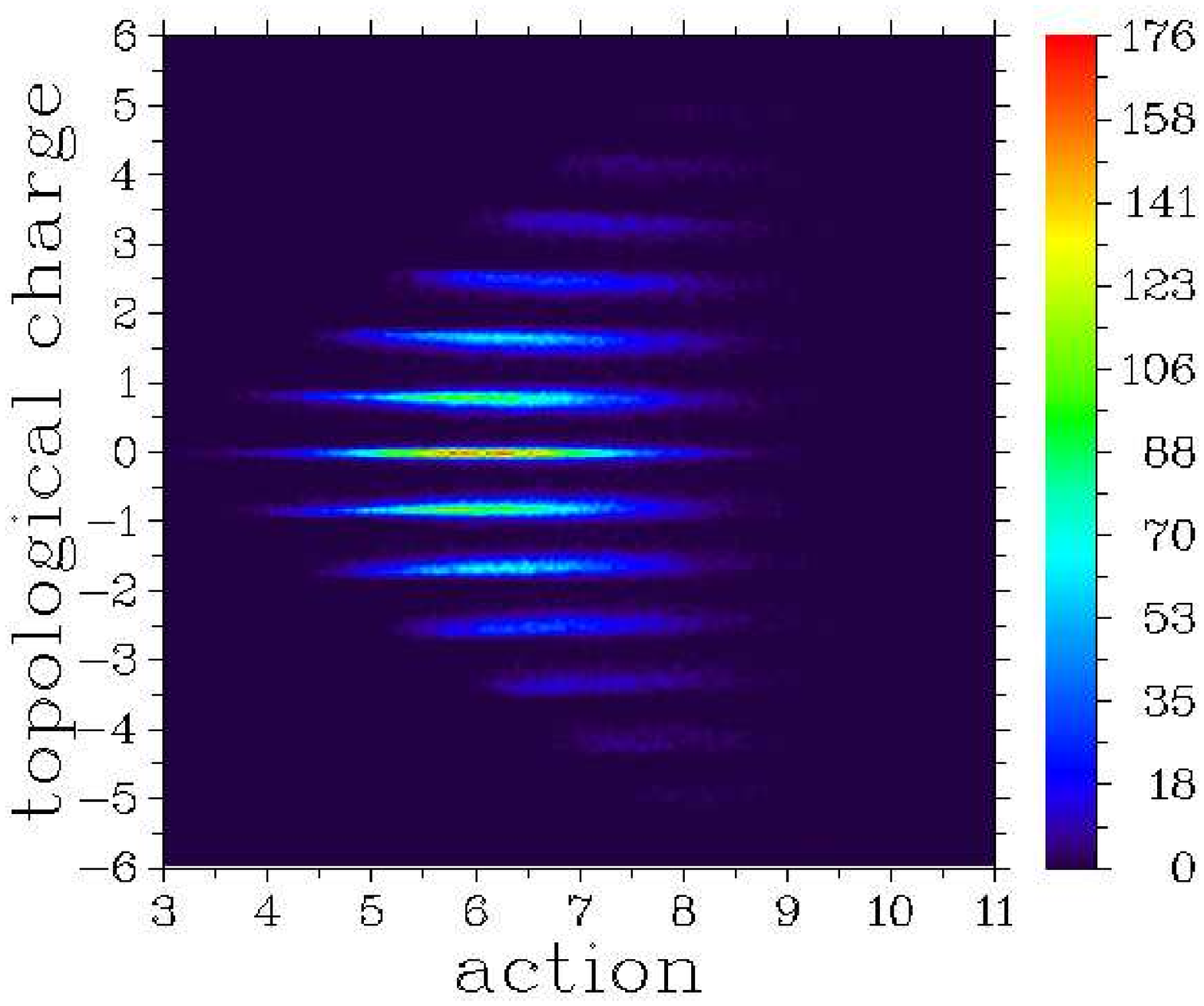} \hspace{0.5cm}
\includegraphics[width=0.42\linewidth]{lump_ratio_pap.eps} 
\caption{Left: action versus topological charge in a 2d-histogram
(the colour-code gives the counts).
Right: Fraction of the total action and of the total topological
charge carried by the leading, {\it action-based} clusters up to ordinal
number $n$ (see the main text for a proper definition of the cluster
construction and enumeration).  }
\label{fig:7_action_charge_scatterplot}
\end{figure}

\vskip 0.3cm
The left panel of Figure~\ref{fig:7_action_charge_scatterplot}
shows a scatterplot of the {\it total}
action versus the {\it total} topological charge. An ensemble of $4000$
independent configurations prepared at $\beta =2.3$ on a $10^4$ lattice has
contributed to this result. We observe that, while the topological
charge $Q$ is quantised to a good extent as expected, the normalised action
(\ref{eq:in6b}) is widely spread in the scatter plot. This result is strikingly
different from what a dilute gas pattern of instantons and antiinstantons 
would show: in
that case, the action would be quantised as well (see~(\ref{eq:in6})).

\vskip 0.3cm
In the further analysis of the space-time structure, we have to identify
the ``lumps'', i.e., clusters in a given configuration. We will employ 
different definitions of the clusters and compare the cluster properties 
arising in either case.

\goodbreak

\vskip 0.3cm
{\it Action-based clusters:}

\begin{itemize}
\item[(i)] At first, we scan the lattice for the maximal value
$s_\mathrm{max}$ of the action density. 
Let this maximal value occur at point $x_0$.

\item[(ii)] Then, we label all points $x_k$ which form a cluster connected 
to $x_0$ and with an action density $s(x_k) \ge t \times s_\mathrm{max}$.
All ``connected'' sites are connected to each other (directly or indirectly) 
by links. The threshold parameter $t$ can be chosen $0 \le t <1$. 
The set of all these points $x_k$ defines the first cluster $C_1$ of the given 
configuration. This cluster gets the ordinal number $n=1$. In later cycles 
of this cluster algorithm (see below) 
further clusters $C_n$ with 
$n=2, 3, ...$ are defined. We then sum over the action density, 
$a_n = \sum_{x \in C_n} s(x)$, and the topological charge density,
$q_n = \sum_{x \in C_n} q(x)$, of all points which have been identified
previously as part of the cluster $C_n$. Apart from action and topological 
charge, each cluster is characterised by $v_n = \sum_{x \in C_n} 1$, the 
number of points (cluster volume).

\item[(iii)] We go back to step (i) and repeat the search for the maximal
action density, but this time points which have previously been identified 
to be part of a cluster are ignored.

\end{itemize}

After the cycle (i) to (iii) was repeated until all lattice points have been
assigned to a cluster, we sort all the clusters according to their action $a_n$ 
in descending order~\footnote{Therefore, these clusters
are called {\it action-based}.} and store for the $10$ biggest clusters $a_n$, 
$q_n$ and $v_n$ for further analysis.

\begin{figure}
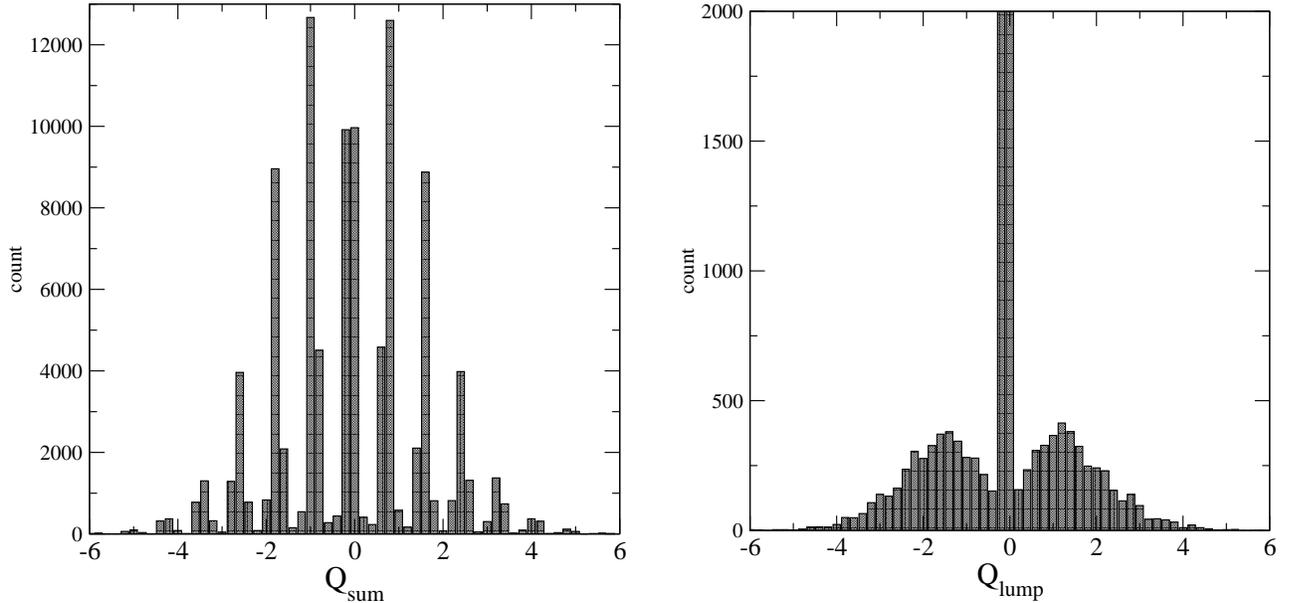

\includegraphics[height=8cm]{qa0_pap.eps} \hspace{0.5cm}
\includegraphics[height=8cm]{q_abs_indv_pap.eps}
\caption{Left: histogram with respect to the summed
topological charge $q_n$ of charge-based clusters.
Right: histogram of $q_n$ provided by the individual charge-based
clusters. Note that the central peak, which reaches as high as $16000$ counts
was capped at $2000$ counts to fit into the figure.}
\label{fig:8_action-based_vs_charge-based}
\end{figure}
To give the reader an impression,
a typical result from a single lattice configuration for a $10^4$ lattice 
prepared at $\beta =2.3$ before it underwent constrained cooling, is listed 
in the left half of Table~\ref{tab:1}.
Since the clusters
are ordered and indexed according to the action, say, by $n=1 \ldots 10$,
we can ask how much of the total action and how much of the total topological
charge is carried by a small number of leading clusters.
To answer these questions, we define the cumulative quantities
\bea
A_n & = & \sum _{i=1}^n a_i \; , \nonumber \\
Q_n & = & \sum _{i=1}^n q_i \; , \\
V_n & = & \sum _{i=1}^n v_i \; . \nonumber
\label{eq:in13}
\ena
Note that all three quantities implicitly depend on the threshold $t$ parameter 
via the cluster definition. The right panel of
Figure~\ref{fig:7_action_charge_scatterplot} shows that, with the above
definition of the action-based clusters, a few of them already account 
for a good deal of both, action and topological charge of the whole 
configuration.

\vskip 0.3cm
In order to get further insights into the texture, we also defined the
clusters using the topological charge density rather than the
action density. Those clusters
are obtained as follows.

\vskip 0.3cm
{\it Charge-based clusters:}

\begin{itemize}
\item[(i)] At first, we search for the maximal value of
$\vert q(x) \vert $, where $q(x)$
is the topological charge density defined in (\ref{eq:in5}).
Assume that this maximal value occurs at point $x_0$.
We then define the ``sign'' of the cluster to be defined first as
$\sigma = \hbox{sign} (q(x_0))$.  

\item[(ii)] Then, we search all points $x_k$ which are connected to $x_0$
and which satisfy $\sigma \, q(x) \ge t \, \vert q(x_0) \vert$,
where $t$ is the threshold parameter.
The set of all these points defines the first {\it charge-based} cluster $C_1$
(with ordinal number $n=1$) of the given configuration.
In later cycles of this cluster algorithm (see below)
further clusters $C_n$ with $n=2, 3, ...$ are defined. 
As in the case 
of action-based clusters, the action
$a_n = \sum_{x \in C_n} s(x)$, the topological charge
$q_n = \sum_{x \in C_n} q(x)$ and the volume
$v_n = \sum_{x \in C_n} 1$ are assigned to the cluster.

\item[(iii)] We go back to step (i) and repeat the scan for
the maximal value of $\vert q(x) \vert $ ignoring those points
which already are part of a cluster.

\end{itemize}

After the cycle (i) to (iii) was repeated until all lattice points have been
assigned to a cluster,
this time we sort all the 
clusters according to the modulus of the topological charge $q_n$ in descending 
order and store the data of the first $10$ clusters.

\vskip 0.3cm
A sample result of the cluster analysis with a threshold $t=10\%$ 
applied to a single configuration on a $10^4$ lattice, Monte-Carlo generated 
at $\beta=2.3$ and then constrained-cooled, can be found in the right
half of Table~\ref{tab:1}.
We then sum up the topological charges of the $10$ leading charge-based
clusters. By definition, summing up the charge of all clusters
would yield the total topological charge $Q$ of the lattice.
We find, however, that -- in particular if the threshold $t$ is chosen smaller 
than $20\%$ -- the sum of the leading $10$ lumps already gives a very good 
approximation to the total topological charge, i.e.,
$Q_{10} \approx Q$. We emphasise the observation that applying a lower threshold 
parameter $t$ leads to a more rapid convergence of the sum for the cumulative 
action {\it and} cumulative topological charge for the given configuration. 

The left panel of Figure~\ref{fig:8_action-based_vs_charge-based}
(left panel) shows the abundance of the cumulative topological charges 
$Q_{\rm sum}=Q_{10}$ 
for charged-based clusters using the minimal threshold $t=0$ for the cluster
analysis of $10^4$ lattices prepared at $\beta = 2.3$.
For this result, $4000$ independent configurations have been analysed
for their cluster structure. The lattice configurations underwent $1000$ sweeps of 
constrained cooling. For small values of $Q_{10}$, the cumulative topological 
charge peaks at integer values while for larger values
a gradual shift to smaller, non-integer values is observed.
We interpret these findings as follows: for small values of $Q_{10}$ the
lump texture is quite smooth allowing for an accurate estimate of the 
(integer) topological charge
using the naive clover type of definition. For larger values of $Q_{10}$,
the configurations are more rough and call for an improved definition
of the topological charge. 
With this caveat, connected clusters of action
with charges up to $Q_{10} \le 6$ have been observed.

\vskip 0.3cm
\begin{figure}
\includegraphics[width=0.325\textwidth]{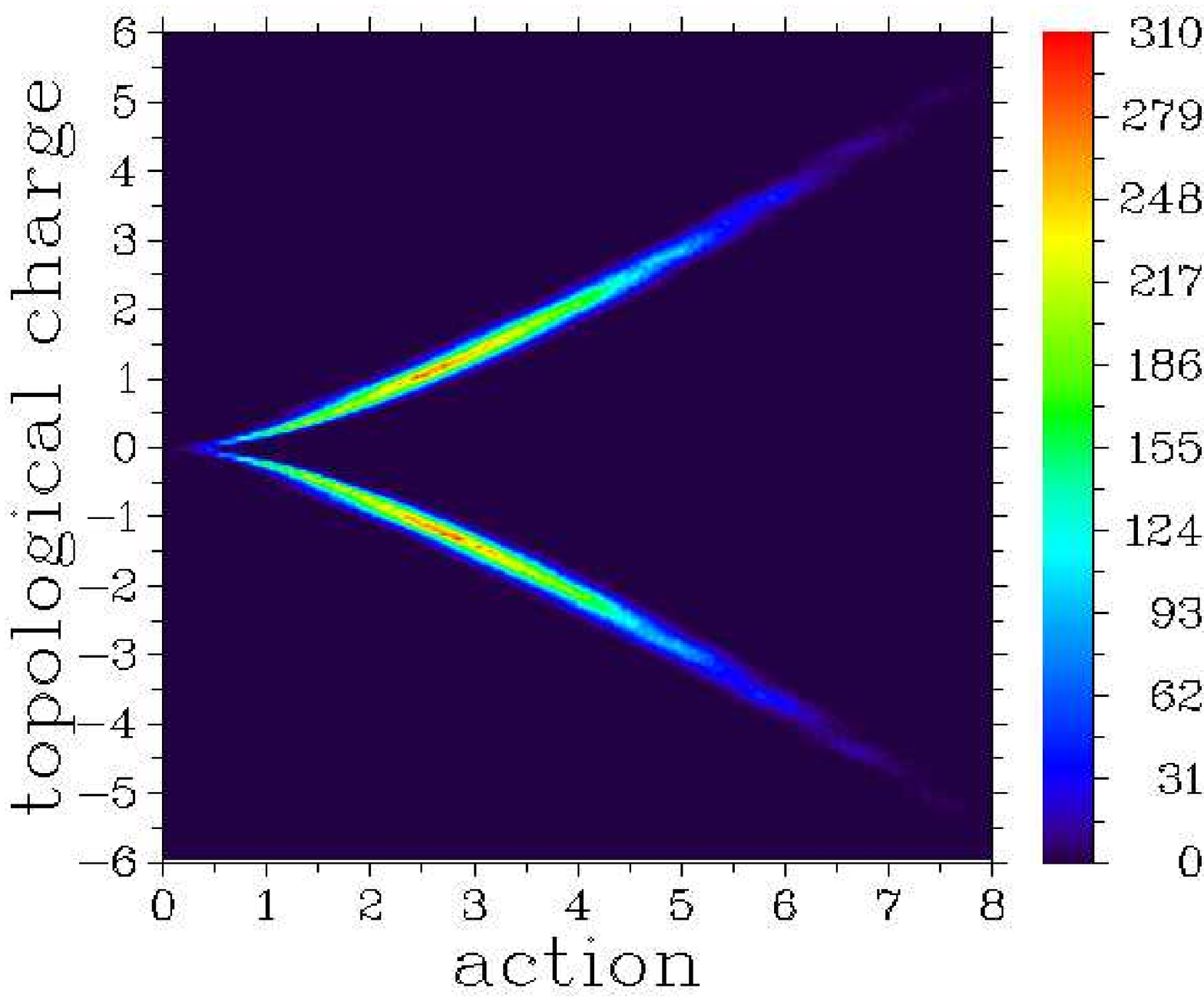}
\includegraphics[width=0.325\textwidth]{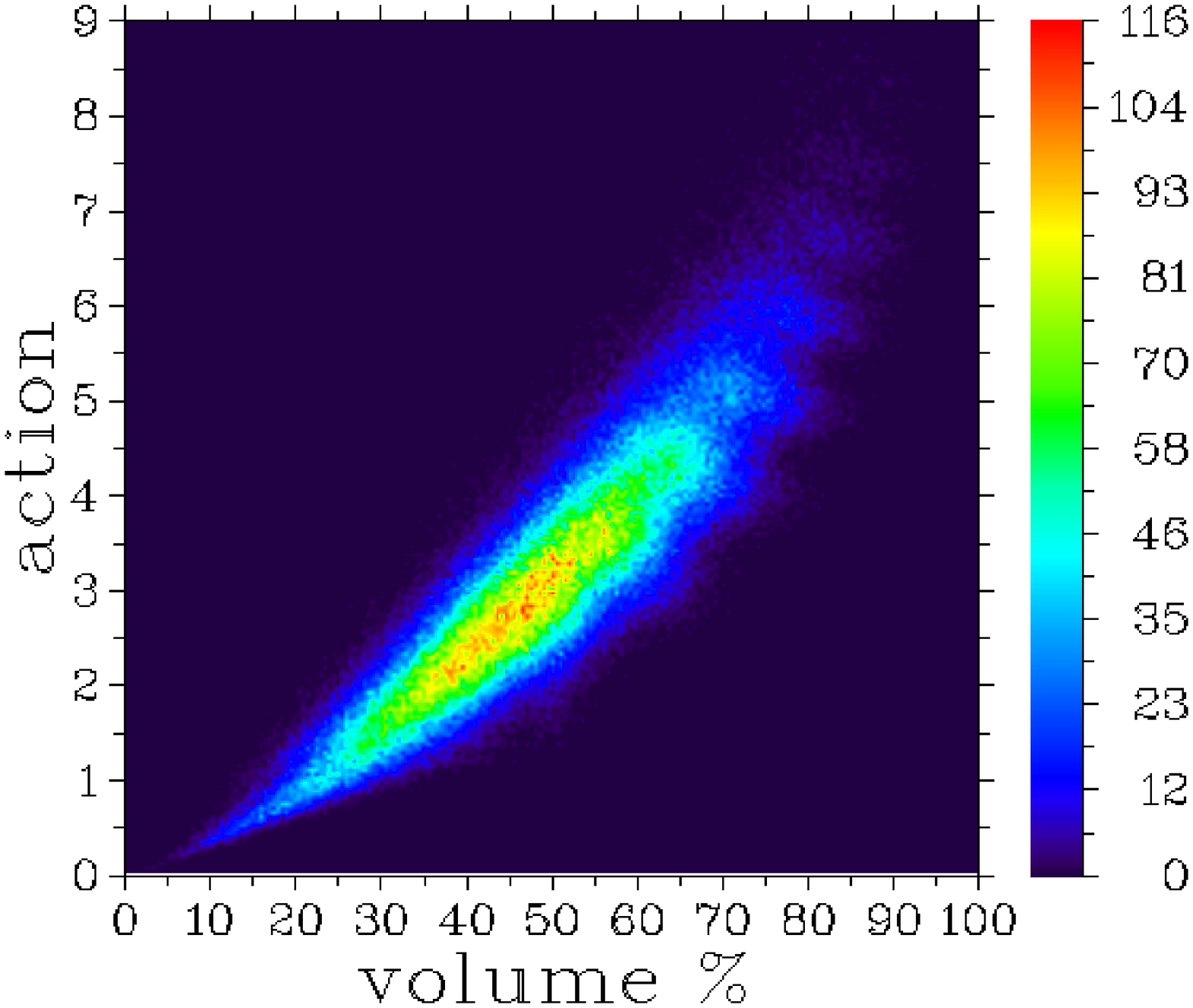}
\includegraphics[width=0.325\textwidth]{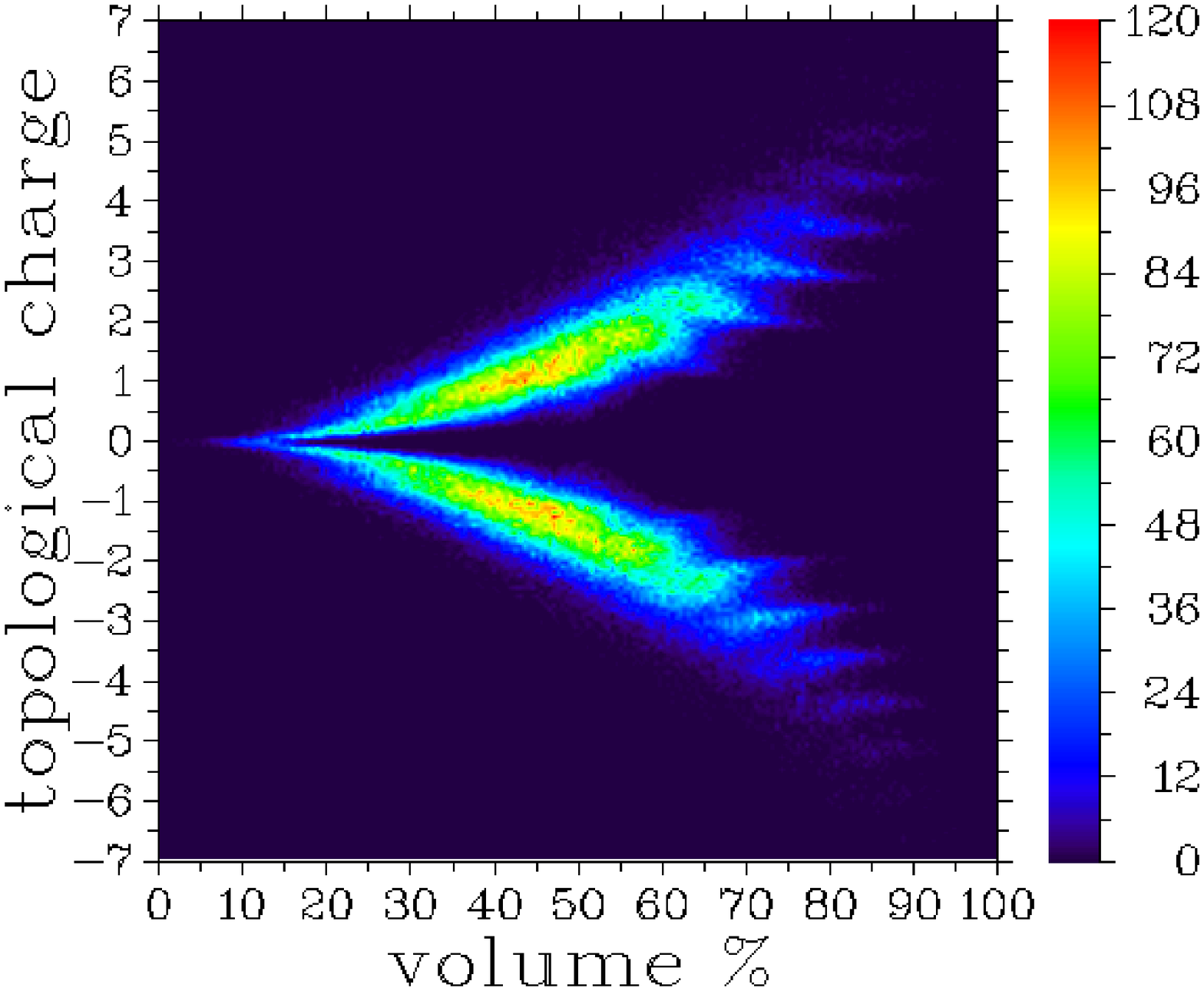}
\caption{\label{fig:sc} Scatter plots involving the volume
$v_n$, the action $a_n$ and the topological charge $q_n$ of all
individual charge-based clusters. }
\label{fig:9_scatter_AVQ}
\end{figure}
In order to investigate whether the topological charge $Q_{\rm lum}=q_n$
of the individual {\it charge-based} clusters is quantised, we have also
calculated their abundance. The result is shown in the right panel of
Figure~\ref{fig:8_action-based_vs_charge-based}. The cluster decomposition
was performed using a threshold parameter $t=0$.
We observe a sharp peak around $q_n \approx 0$.
This peak originates from a huge number of isolated micro-clusters
(fluctuations) which contribute very little to the topological charge. 
More striking is the fact
that the distribution is smooth for larger values of $q_n$ with a maximum
around $ \approx 2$. No sign of an integer quantisation is visible as one would
expect from a dilute gas of instantons. The charge-based clusters carry a broad 
spectrum of topological charges. Only the sum of their individual contributions
is quantised.

\vskip 0.3cm
In Figure~\ref{fig:9_scatter_AVQ}, we finally present scatter plots of data
describing individual charge-based clusters with minimal threshold $t=0$,
involving their volume $v_n$, action $a_n$ and topological charge $q_n$.
Each point in these figures corresponds to one cluster. There is a close
correlation and monotonic relation between action and topological charge: 
the higher the action, the higher is the modulus of charge. This relation 
is not far from saturation, $\vert q_n \vert \lesssim \frac{3}{4} a_n$. 
Nevertheless, the space-time distribution of action and topological charge 
is not identical. Note that, in contrast to an instanton or meron gas 
configuration, the topological charge varies smoothly with the volume or the 
action of the cluster. We also observe a smooth correlation between the volume 
of a cluster and its charge and action, respectively. There are bounds
$a_n \le 10 v_n/N^4$ and $\vert q_n \vert \le 7 v_n/N^4$.

\subsection{Selfduality and inverse participation ratios \label{sec:ipr} }

\begin{figure}
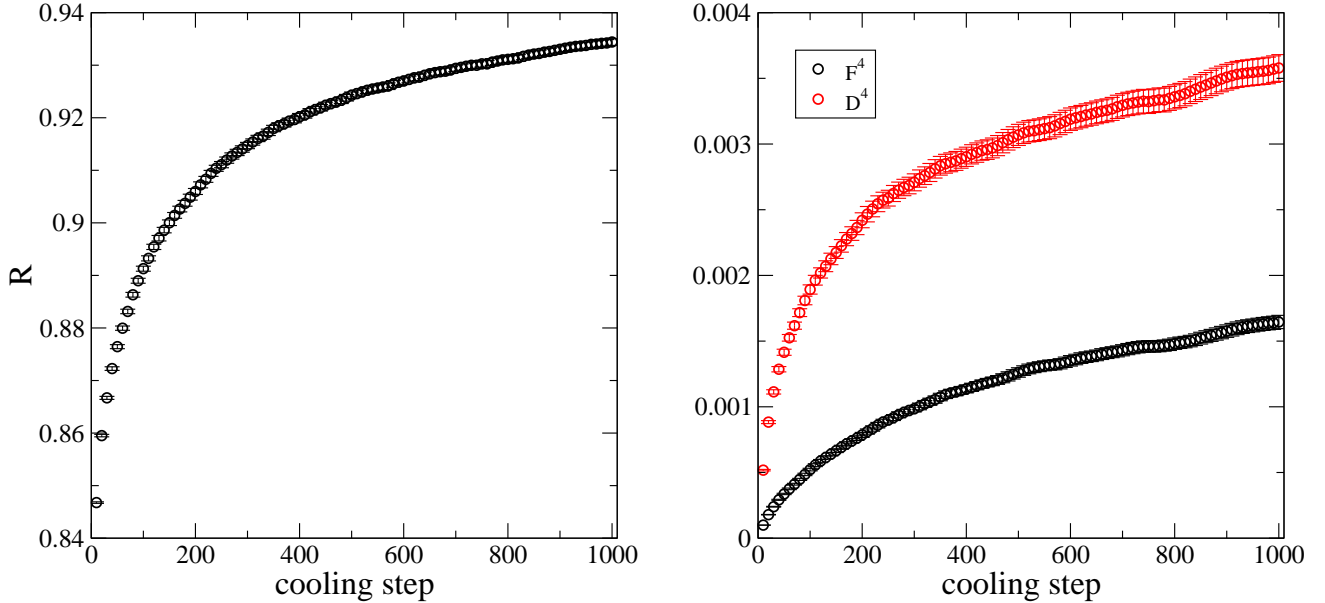

\includegraphics[height=8cm]{dual_pap.eps} \hspace{0.5cm}
\includegraphics[height=8cm]{ipr_pap.eps}
\caption{Left: the selfduality measure $R$. Right: the inverse participation
ratios $F^4$ and $D^4$.
All quantities are shown as functions of the cooling sweep.}
\label{fig:10_selfduality_IPR}
\end{figure}
In the following, we will analyse to which extent the clusters are
selfdual or anti-selfdual  and which portion of space-time they do occupy.
For this purpose, we introduce for each lattice configurations two normalised
densities, ``unit'' vectors derived from $s(x)$ and $q(x)$:
\bea
{\cal F}_x [U] &=& \frac{s(x)}{\sqrt{\sum _x s(x)^2}} \, , \hbo
\sum _x {\cal F}_x {\cal F}_x =1 \; ,
\label{eq:in15} \\
{\cal D}_x [U] &=& \frac{q(x)}{\sqrt{\sum _x q(x)^2}} \, , \hbo
\sum _x {\cal D}_x {\cal D}_x =1 \; .
\label{eq:in16}
\ena
At first we notice that the ensemble expectation value
\be
\Bigl\langle \sum _x \, {\cal F}_x[U] {\cal D}_x[U] \, \Bigr\rangle
\en
vanishes upon the integration over the link fields.
In order to find out whether a high field strength is linked to
a high modulus of the topological charge density, we consider
the expectation value $R$,
\be
0 \; \le \; R \; = \;
\Biggl\langle \sum _x \, {\cal F}_x[U] \, \Bigl\vert {\cal D}_x[U] \Bigl\vert
\, \Biggr\rangle \; \le \; 1 \; .
\label{eq:in17}
\en
Note that $R=1$ corresponds to a perfect uniform alignment between
the modulus of the topological charge density and the field strength.
Let us also point out that the overall scale of the colour electric
and colour magnetic 
action density drops out from the definitions for $R$, ${\cal F}$
and ${\cal D}$ implying that not only regions
of large action and charge density 
contribute. Also ``small'', aligned fluctuations can make a sizable 
contribution to these quantities.

\vskip 0.3cm
We have studied $R$ for a $16^4$ lattice and for $\beta = 2.4$. The
result is shown in the left panel of Figure~\ref{fig:10_selfduality_IPR}
as a function of the cooling sweep. We find that $R$ gradually increases
with cooling reaching values as large as $0.9344(6)$ after $1000$ cooling
sweeps. This finding indicates that the constraints, imposed upon cooling,
do not hamper 
to reach a high degree of selfduality of the final cooled configuration.

\vskip 0.3cm
\begin{table}
\centering
\begin{tabular}{c|cccc}
 & $14^4$ & $16^4$ & $18^4$ & $20^4$ \\ \hline
$R$     & $0.9244(1)$ & $0.9147(5)$ & $0.9379(5)$ & $0.9432(4)$ \\
$F^4$   & $2.1(1) \times 10^{-3} $ & $1.64(5) \times 10^{-3} $ 
        & $1.18(3) \times 10^{-3} $ & $0.80(1) \times 10^{-3} $ \\
$D^4$   & $0.50(2) \times 10^{-2} $ & $0.36(1) \times 10^{-2} $
        & $0.248(7) \times 10^{-2} $ & $0.160(3) \times 10^{-2} $ \\
\end{tabular}
\caption{\label{tab:2} The selfduality measure $R$ and
the inverse participation ratios for the field strength and the modulus
of the topological charge density for $1000$ constrained cooling sweeps. }
\end{table}
Furthermore, we studied the inverse participation ratios (IPRs) of the action 
density and the modulus of the topological density:
\be
F^4 \; := \; \Biggl\langle \sum _x {\cal F}_x^4 \Biggr\rangle \; , \hbo
D^4 \; := \; \Biggl\langle \sum _x {\cal D}_x^4 \Biggr\rangle \; ,
\label{eq:in18}
\en
If the configurations would be extremely localised, i.e.,
${\cal F}_{x_0}$ would be not vanishing only at one site,
we would find $F^4=1$. On the other hand, if ${\cal F}$ and ${\cal D}$
are smeared out over a sub-manifold of dimension $d$, we would obtain
(with $V=N^4$ the lattice volume)
$$
F^4 \; = \; \frac{1}{N^d} \to 0 \; .
$$
Our result for $F^4$ and $D^4$ for a fixed number $N$
is shown in the right panel of Figure~\ref{fig:10_selfduality_IPR}.
We find that these quantities are increasing during cooling.
This indicates an increase of the localisation of the configurations
during cooling as expected. We also point out that the IPR for
topological charge density, i.e., $D^4$, is roughly twice as big as
$F^4$. This finding signals that the topological charge density is more
localised than the action density.
This could mean that the clusters of
action possess one or more selfdual centres (giving rise to the localised
topological charge density) and an halo of action made from more generic
(not necessarily selfdual) fields.

\vskip 0.3cm
\begin{figure}
\begin{center}
\includegraphics[height=8cm]{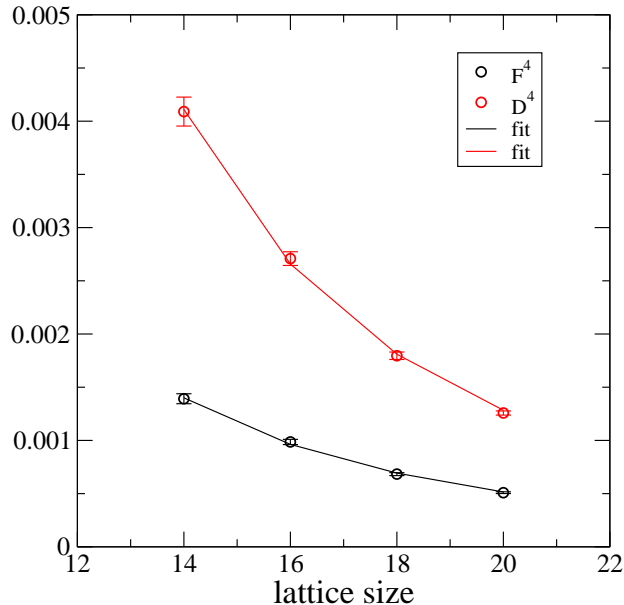}
\end{center}
\caption{The inverse participation ratios 
$\langle {\cal F}^4 \rangle$ and $\langle {\cal D}^4 \rangle$  
after $1000$ constrained cooling sweeps as a function of the linear lattice 
size $N$ ($N^4$ being the lattice volume) for 400 configurations of each 
size prepared at $\beta=2.4$. }
\label{fig:11_volume_dependence}
\end{figure}
In order to study the effective dimension $d$ of the cluster
structure, the IPRs $F^4$ and $D^4$ (and also  $R$) were calculated
for configurations with the lattice sizes $14^4$, $16^4$, $18^4$
and $20^4$, all prepared at $\beta =2.4$. 
The results obtained from an average over $400$
independent configuration are summarised in Table~\ref{tab:2} and
in Figure~\ref{fig:11_volume_dependence}. The results are fitted by
\be
F^4 \approx 1.9(1) \, \times \, N^{-2.6(1)}, \hbo
D^4 \approx 13(1) \, \times \, N^{-3.0(1)}.
\label{eq:in19}
\en
Note that not only the absolute values for the  IPRs
$F^4$ and $D^4$ are quite different as already noticed, but also
the exponents are slightly different.
At the same time, we observed a strong correlation between regions of large 
action density and of large topological charge density (brought to light by
$R$). 
This indicates that there is a subtle difference in the arrangements of
action density and topological charge density which are hardly visible by
the naked eye (see Figure~\ref{fig:6_action_and_topcharge_visual}).
Let us also stress that the data fitted by (\ref{eq:in19}) are obtained for 
a {\it fixed} number of cooling steps (i.e. $1000$ sweeps). These figures 
might still change if the number of cooling sweeps is further increased. 
A systematic study of this effect is left to future work.

\section{Spontaneous chiral symmetry breaking}
\label{sec:SSB}

\begin{figure}
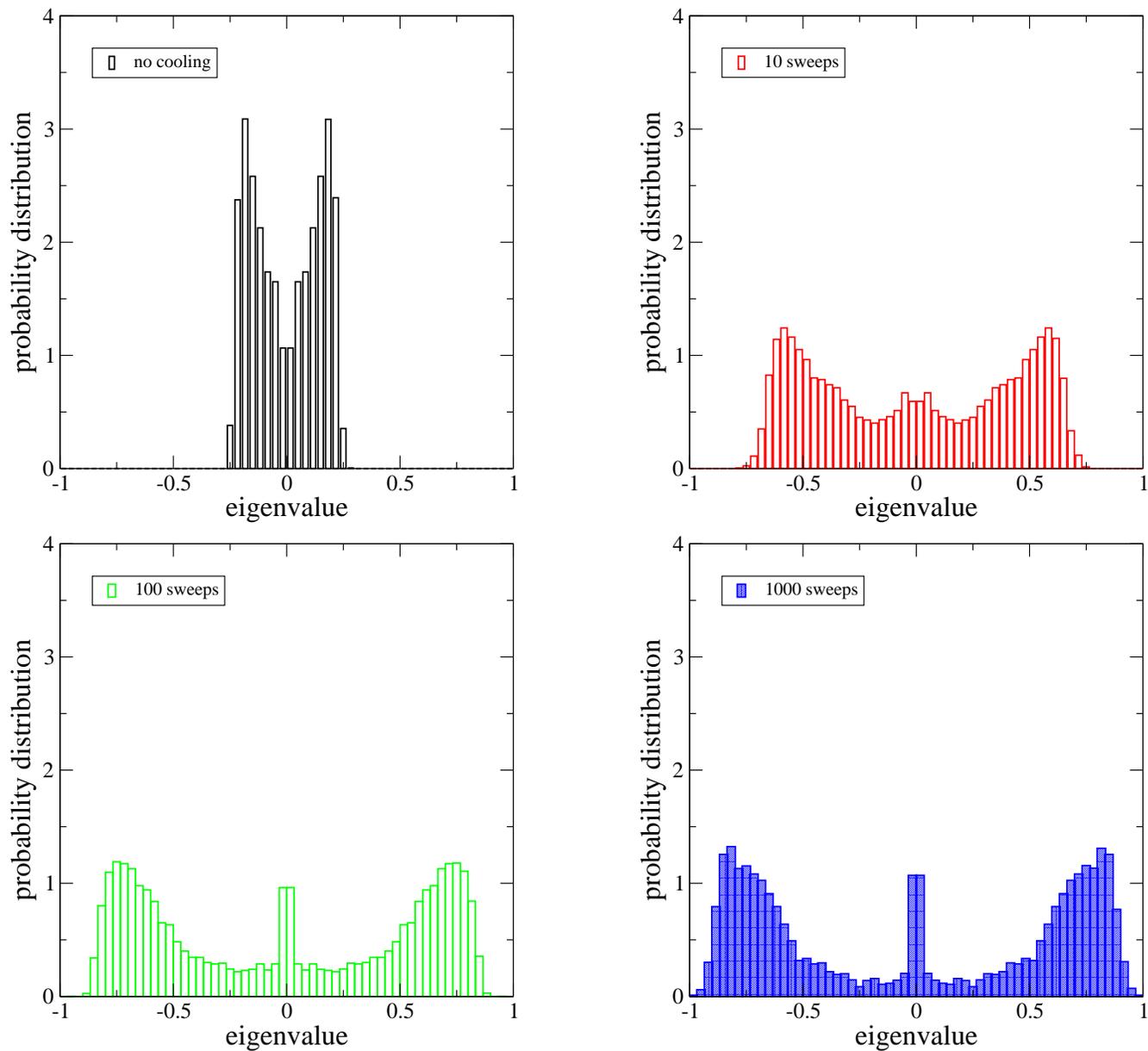

\begin{tabular}{lcl}
\includegraphics[height=8cm]{stagg_0_pap.eps} & \hspace{1cm} &
\includegraphics[height=8cm]{stagg_10_pap.eps} \\
\includegraphics[height=8cm]{stagg_100_pap.eps} & \hspace{1cm} &
\includegraphics[height=8cm]{stagg_1000_pap.eps}
\end{tabular}
\caption{The spectral density of the staggered Dirac 
operator as a function of its eigenvalues for several 
numbers of constrained cooling sweeps. }
\label{fig:12_fermion_spectrum_constrained_cooling}
\end{figure}
\begin{figure}
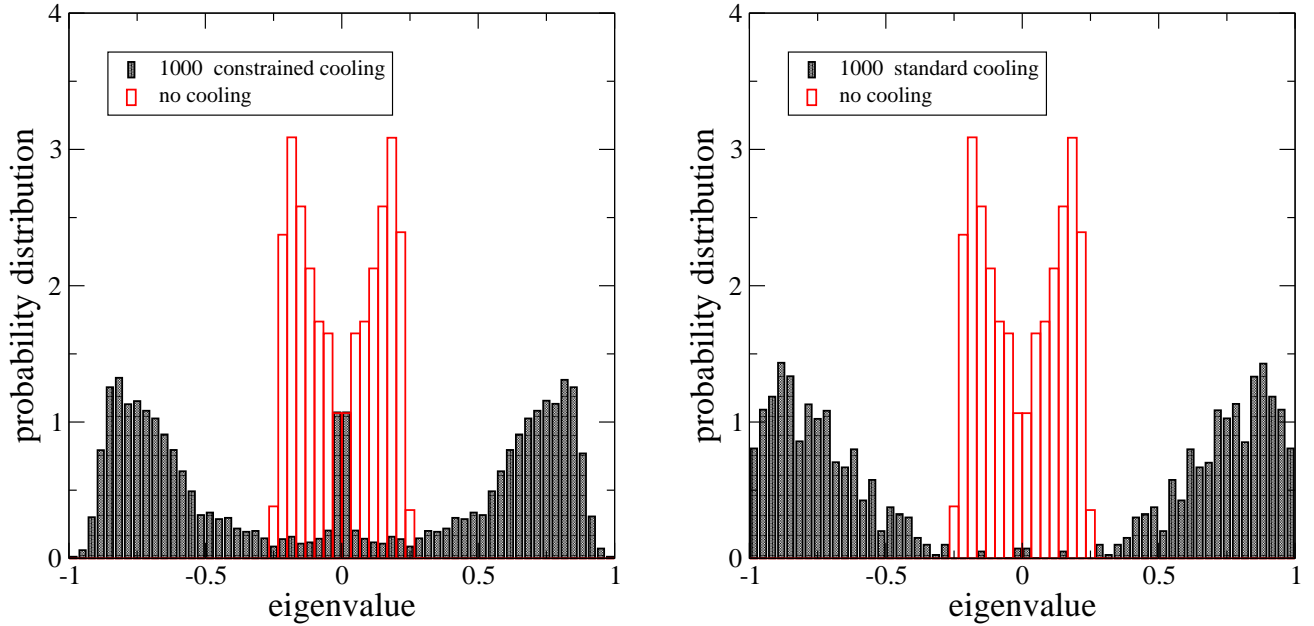

\includegraphics[height=8.3cm]{stagg_dis2_pap.eps} \hspace{.7cm}
\includegraphics[height=8.3cm]{stagg_stan_dis_pap.eps} 
\caption{The spectral density of the staggered Dirac 
operator as a function of its eigenvalues after 1000 sweeps of
cooling. Left: for constrained cooling; right: for standard cooling. }
\label{fig:13_fermion_spectrum_comparison}
\end{figure}

Besides confinement, the spontaneous breakdown of chiral
symmetry is another outstanding feature of QCD
which shapes the hadron spectrum and therefore the structure of matter.
The crucial test
for spontaneous chiral symmetry breaking is a non-vanishing quark condensate
$\langle \bar{q}q \rangle$. It is related
to the eigenvalue spectrum $\rho(\lambda)$ of the fermion operator
$M[U]$ by virtue of the Banks-Casher relation:
\be
\langle \bar{q}q \rangle \; = \; \pi \, \rho(0) \; .
\en
For simplicity, we adopt staggered fermions which avoid fermion doubling
while keeping a subset 
of chiral symmetry intact. The massless Dirac operator is
\be
M[U] \; = \; \sum _{\mu =1}^4 \eta _\mu (x) \, \Bigl[ U_\mu(x)
\, \delta _{x + \mu, y} \; - \; U^\dagger _\mu (x-\mu)
\, \delta _{x -\mu, y} \, \Bigr] \; ,
\label{eq:f1}
\en
where the phase factors are given by
$$
 \eta _\mu(x) \; = \; (-1)^{x_1 + \ldots + x_{\mu -1} } \; .
$$
We have numerically estimated the spectrum of the low lying
modes for configurations created on a $10^4$ lattice with $\beta = 2.4$.
For this aim, we have calculated the $50$ modes with $\lambda$ closest to zero,
taking them as entries to a histogram. We then averaged the histograms over 
$400$ independent configurations. The result for un-cooled configurations is
shown in the upper left picture in
Figure~\ref{fig:12_fermion_spectrum_constrained_cooling}.
As expected, the eigenvalue spectrum is non-vanishing at zero eigenvalues. 
We subsequently studied the eigenvalues spectrum after
$10$, $100$ and $1000$ sweeps of constrained cooling.
It turns out that cooling pushes the eigenvalues away from zero 
while a group of eigenvalues remains forming a peak near zero eigenvalue. 
It is striking that number of near-zero modes firstly decreases, but that 
the full strength of the spectral density near zero seems to be regained 
asymptotically. We finally compare the spectrum after 1000 sweeps of our 
new constrained cooling technique with that from standard cooling
(see figure~\ref{fig:13_fermion_spectrum_comparison}).
While the near zero-modes are protected throughout the constrained cooling
sweeps, the near-zero modes have practically disappeared after 1000 sweeps of
standard cooling.

\section{Conclusions \label{sec:conclusions} }

Standard cooling finally approaches classical Yang-Mills configurations
which lack important low energy features of the quantized theory such
as confinement. As a tool for analysis, standard cooling has to be stopped 
by an ad hoc criterion. Although it might be possible by this procedure 
to generate an ensemble of configurations which grasp the essence of low 
energy Yang-Mills theory, the persuasiveness of the approach is flawed
since the properties of the configurations might still depend on the
stopping parameters.

\vskip 0.3cm
{\it Constrained cooling } in the spirit firstly proposed
in~\cite{Langfeld:2009es}  and detailed in the present paper removes this
deficiency. Here, as well as in~\cite{Langfeld:2009es}, the respective physical
constraint makes it unnecessary  
to monitor and stop the cooling process at all. 
Instead, the physical constraints ensure that the configurations hardly change 
after many cooling sweeps and that, at the same time, the emerging (semi-)
classical configurations retain the important IR properties. A natural choice
for such a  constraint is to retain the asymptotic form of the static
quark-antiquark potential. In this paper, we achieved this by demanding that
every Polyakov line on the  lattice, i.e., through each lattice site and in
every direction, does not change  during the cooling process. Despite of the
huge amount of constraints, we find that  smooth (semi-) classical
configurations with very low action emerge  
(see the discussion in subsection~\ref{sec:ConstrainedCooling} which quantifies
this statement). Once the confining property of the (semi-) classical
configurations 
is established, we find that other important low energy properties are conserved
as  well: the topological susceptibility is largely unaffected by constrained
cooling, and the investigation of the eigenvalue spectrum of the staggered
Dirac operator  indicates that these configurations provide spontaneous chiral
symmetry breaking as well. 

\vskip 0.3cm
A thorough analysis of the space-time structure of the configuration emerging 
from {\it constrained cooling} has been performed. It turns out that these 
configurations can be described by connected clusters 
which are (anti-) selfdual to a good extent (see subsection~\ref{sec:ipr}
for quantitative details). We find that these objects are genuinely different  
from instantons: while the total topological charge of each configuration is 
quantised in integers (as it should be), the topological charge of individual 
clusters takes values which are continuously distributed.
Data showing the topological charge of {\it individual} clusters versus their 
action, see figure~\ref{fig:9_scatter_AVQ}, seem to fall on a well-defined
curve in the action-topological charge histogram (rather than populating
isolated islands as it would be the case for instanton ensembles).

\vskip 0.3cm
To explore these many observables reported here with a high-level statistics,
these first results have been obtained by using an extremely simple lattice 
set-up. In view of our findings, it is certainly worthwhile to further study 
the configurations exposed by {\it constrained cooling} employing an improved 
lattice action, an improved definition of the topological charge density and 
a more elaborate Dirac operator with maximal chiral symmetry.
We also expect that extending the investigations to finite temperatures will 
provide valuable insight in the nature of the deconfinement phase transition 
and possibly in some features of the quark gluon plasma close to $T_c$.
This, as well as an extension to $SU(3)$, is planned for future work.

\bigskip 
{\bf Acknowledgments: } 
The calculations have been carried out within the DiRAC framework 
at the {\tt High Performance Computing Centre} at University of Plymouth. 
We are indebted to the staff for support. This work is a project of the 
UKQCD collaboration. DiRAC is supported by STFC. 

Both authors acknowledge discussions with Biagio Lucini and Derek Leinweber.
E.-M. I. is grateful for a temporal position at the Faculty of Physics
of the University of Bielefeld he held during the first phase of this project.
He is obliged to his collaborators in the QCDSF collaboration, in particular
Gerrit Schierholz and Volker Weinberg, in the collaboration between Humboldt 
University and ITEP, in particular Michael M\"uller-Preussker and Boris
Martemyanov, and recently at the University of Regensburg, in particular 
Falk Bruckmann and Florian Gruber, for many discussions concerning the 
topological structure of QCD and its description.

\appendix
\section{Solving the constraint \label{sec:appA} }

Using (\ref{eq:12}), the task is to determine the Lagrange multiplier such that
the constraint in (\ref{eq:10}) is satisfied. We firstly note that
$$
{\cal N}^2 \; = \; \det(B_\mu) \, + \,  \lambda _\mu \, \tr (B_\mu^\dagger
{\cal P_\mu }) \, + \, \lambda _\mu ^2 ,
$$
and squaring the constraint equation (\ref{eq:10}) leads to a
quadratic equation for $\lambda _\mu $:
$$
\lambda _\mu ^2 \; + \;  \lambda _\mu \, \tr(B_\mu^\dagger {\cal P}_\mu )
\; + \;
\frac{ \frac{1}{4} \tr ^2 (B_\mu^\dagger {\cal P}_\mu ) -
\mathrm{det} (B_\mu) \, \frac{1}{4} \tr^2 (U_\mu {\cal P}_\mu) }{
1 - \frac{1}{4} \tr^2 (U_\mu {\cal P}_\mu) } \; = \; 0 .
$$
Since both $U_\mu ,\,  {\cal P}_\mu \, \in \, $SU(2), we point out out that
$$
1 - \frac{1}{4} \tr^2 (U_\mu {\cal P}_\mu)  \; \ge \; 0 \; .
$$
We  find two solutions:
$$
\lambda _\mu ^\pm \; = \; - \frac{1}{2} \,
\tr (B_\mu^\dagger {\cal P}_\mu ) \; \pm \; \frac{1}{2}
\tr (U_\mu {\cal P}_\mu ) \; \sqrt{
\frac{ \mathrm{det} (B_\mu) \, - \, \frac{1}{4}
\tr^2 (B^\dagger_\mu {\cal P}_\mu) }{ 1 - \frac{1}{4} \tr^2 (U_\mu {\cal P}_\mu) }} .
$$
One of the solution is spurious, since we have squared the constraint
equation to derive them. Indeed, calculating
$$
\frac{\cal N}{2} \, \tr(U_\mu ^c  {\cal P} _\mu ) \; = \;
\frac{1}{2} \tr \Bigl[(B_\mu^\dagger + \lambda _\mu ^\pm \, {\cal P}^\dagger _\mu )
{\cal P_\mu }\Bigr] \; = \; \pm \; \frac{1}{2}
\tr (U_\mu {\cal P}_\mu ) \; \sqrt{
\frac{ \mathrm{det} (B_\mu) \, - \, \frac{1}{4}
\tr^2 (B^\dagger_\mu {\cal P}_\mu) }{ 1 - \frac{1}{4} \tr^2 (U_\mu {\cal P}_\mu) }}
$$
rules out the solution $ \lambda _\mu ^-$ since ${\cal N}>0$. We also find
the normalisation factor to be
$$
{\cal N} \; = \; \sqrt{ \frac{ \mathrm{det} (B_\mu) \, - \, \frac{1}{4}
\tr^2 (B^\dagger_\mu {\cal P}_\mu) }{ 1 - \frac{1}{4} \tr^2
(U_\mu {\cal P}_\mu) }} .
$$


\begin{thebibliography}{sch90}

\bibitem{Callan:1977gz}
     C.~G.~Callan Jr., R.~F.~Dashen, and D.~J. Gross,
     Phys.\ Rev.\ {\bf D17} (1978) 2717,

\bibitem{Callan:1978bm}
     C.~G.~Callan Jr., R.~F.~Dashen, and D.~J. Gross,
     Phys.\ Rev.\ {\bf D19} (1979) 1826.

\bibitem{Belavin:1975fg}
     A.~A.~Belavin, A.~M.~Polyakov, A.~S.~Schwartz, and Yu.~S.~Tyupkin,
     Phys.\ Lett.\ {\bf B59} (1975) 85.

\bibitem{Callan:1977qs}
     C.~G.~Callan Jr., R.~F.~Dashen, and D.~J. Gross,
     Phys.\ Lett.\ {\bf B66} (1977) 375.

\bibitem{Teper:1985ek}
     M.~Teper,
     Phys.\ Lett.\ {\bf B171} (1986) 86.

\bibitem{Ilgenfritz:1985dz}
     E.--M.~Ilgenfritz, M.~L.~Laursen, G.~Schierholz, M.~M\"uller-Preussker, and H.~Schiller,
     Nucl.\ Phys.\ {\bf B268} (1986) 693.

\bibitem{de Forcrand:1997sq}
     Ph.~de Forcrand, M.~Garcia~Perez, and I.--O.~Stamatescu,
     Nucl.\ Phys.\ {\bf B499} (1997) 409.
     [\href{http://arxiv.org/abs/hep-lat/9701012}{\tt arXiv:hep-lat/9701012}]]

\bibitem{GarciaPerez:1998ru}
     M.~Garcia Perez, O.~Philipsen, and I.--O.~Stamatescu,
     Nucl.\ Phys.\ {\bf B551} (1999) 293.
     [\href{http://arxiv.org/abs/hep-lat/9812006}{\tt arXiv:hep-lat/9812006}]]

\bibitem{GarciaPerez:1999it}
     M.~Garcia Perez, O.~Philipsen, and I.--O.~Stamatescu,
     Nucl.\ Phys.\ Proc.\ Suppl.\ {\bf 83} (2000) 541.
     [\href{http://arxiv.org/abs/hep-lat/9909029}{\tt arXiv:hep-lat/9909029}]]

\bibitem{Horvath:2002yn}
     I.~Horvath, S.~J.~Dong, T.~Draper, F.~X.~Lee, K.~F.~Liu, H.~B.~Thacker, and J.~B.~Zhang,
     Phys.\ Rev.\ {\bf D67} (2003) 011501.
     [\href{http://arxiv.org/abs/hep-lat/0203027}{\tt arXiv:hep-lat/0203027}]]

\bibitem{Ilgenfritz:2007xu}
     E.-M.~Ilgenfritz, K.~Koller, Y.~Koma, G.~Schierholz, T.~Streuer, and V.~Weinberg,
     Phys.\ Rev.\ {\bf D76} (2007) 034506.
     [\href{http://arxiv.org/abs/0705.0018[hep-lat]}{\tt arXiv:0705.0018[hep-lat]}]
									 
\bibitem{Bruckmann:2005hy}
     F.~Bruckmann and E.-M.~Ilgenfritz, 
     Phys.\ Rev.\ {\bf D72} (2005) 114502.
     [\href{http://arxiv.org/abs/hep-lat/0509020}{\tt arXiv:hep-lat/0509020}]]

\bibitem{Ilgenfritz:2008ia}
     E.-M.~Ilgenfritz, D.~Leinweber, P.~Moran, K.~Koller, G.~Schierholz, and V.~Weinberg,
     Phys.\ Rev.\ {\bf D77} (2008) 074502.  
     [\href{http://arxiv.org/abs/0801.1725[hep-lat]}{\tt arXiv:0801.1725[hep-lat]}]

\bibitem{DeGrand:1996ih}
     T.~A.~DeGrand, A.~Hasenfratz, and D.--c.~Zhu,
     Nucl.\ Phys.\ {\bf B475} (1996) 321.
     [\href{http://arxiv.org/abs/hep-lat/9603015}{\tt arXiv:hep-lat/9603015}]]

\bibitem{DeGrand:1996zb}
     T.~A.~DeGrand, A.~Hasenfratz, and D.--c.~Zhu,
     Nucl.\ Phys.\ {\bf B478} (1996) 349.
     [\href{http://arxiv.org/abs/hep-lat/9604018}{\tt arXiv:hep-lat/9604018}]]

\bibitem{Feurstein:1996cf}
     M.~Feurstein, E.--M.~Ilgenfritz, M.~M\"uller-Preussker, and S.~Thurner,
     Nucl.\ Phys.\ {\bf B511} (1998) 421.
     [\href{http://arxiv.org/abs/hep-lat/9611024}{\tt arXiv:hep-lat/9611024}]]

\bibitem{Feurstein:1997rj}
     M.~Feurstein, E.--M.~Ilgenfritz, H.~Markum, M.~M\"uller-Preussker, and S.~Thurner,
     Nucl.\ Phys.\ Proc.\ Suppl.\ {\bf 63} (1998) 480.
     [\href{http://arxiv.org/abs/hep-lat/9709140}{\tt arXiv:hep-lat/9709140}]]

\bibitem{DeGrand:1997gu}
     T.~A.~DeGrand, A.~Hasenfratz, and T.~G.~Kovacs,
     Nucl.\ Phys.\ {\bf B505} (1997) 417.
     [\href{http://arxiv.org/abs/hep-lat/9705009}{\tt arXiv:hep-lat/9705009}]]

\bibitem{DeGrand:1997sd}
     T.~A.~DeGrand, A.~Hasenfratz, and T.~G.~Kovacs,
     Phys.\ Lett.\ {\bf B420} (1998) 97.
     [\href{http://arxiv.org/abs/hep-lat/9710078}{\tt arXiv:hep-lat/9710078}]]

\bibitem{DeGrand:1997ss}
     T.~A.~DeGrand, A.~Hasenfratz, and T.~G.~Kovacs,
     Nucl.\ Phys.\ {\bf B520} (1998) 301.
     [\href{http://arxiv.org/abs/hep-lat/9711032}{\tt arXiv:hep-lat/9711032}]]

\bibitem{Hasenfratz:1998qk}
     A.~Hasenfratz and C.~Nieter,
     Phys.\ Lett.\ {\bf B439} (1998) 366.
     [\href{http://arxiv.org/abs/hep-lat/9806026}{\tt arXiv:hep-lat/9806026}]]

\bibitem{Hasenfratz:1999ng}
     A.~Hasenfratz,
     Phys.\ Lett.\ {\bf B476} (2000) 188.
     [\href{http://arxiv.org/abs/hep-lat/9912053}{\tt arXiv:hep-lat/9912053}]]

\bibitem{Kovacs:1999an}
     T.~G.~Kovacs,
     Phys.\ Rev.\ {\bf D62} (2000) 034502.
     [\href{http://arxiv.org/abs/hep-lat/9912021}{\tt arXiv:hep-lat/9912021}]]

\bibitem{Kraan:1998pm}
     T.~C.~Kraan and P.~van Baal, 
     Nucl.\ Phys.\ {\bf B533} (1998) 627.
     [\href{http://arxiv.org/abs/hep-th/9805168}{\tt arXiv:hep-th/9805168}]]

\bibitem{Kraan:1998sn}
     T.~C.~Kraan and P.~van Baal,
     Phys.\ Lett.\ {\bf B435} (1998) 389.
     [\href{http://arxiv.org/abs/hep-th/9806034}{\tt arXiv:hep-th/9806034}]]

\bibitem{Lee:1998bb}
     K.--M.~Lee and C.--h.~Lu,
     Phys.\ Rev.\ {\bf D58} (1998) 025011.
     [\href{http://arxiv.org/abs/hep-th/9802108}{\tt arXiv:hep-th/9802108}]]

\bibitem{Gerhold:2006sk}
     Ph.~Gerhold, E.--M.~Ilgenfritz, and M.~M\"uller-Preussker,
     Nucl.\ Phys.\ {\bf B760} (2007) 1.
     [\href{http://arxiv.org/abs/hep-ph/0607315}{\tt arXiv:hep-ph/0607315}]]

\bibitem{Bruckmann:2009pa}
     F.~Bruckmann, E.--M.~Ilgenfritz, B.~V.~Martemyanov, and Bo~Zhang,
     Phys.\ Rev.\ {\bf D81} (2010) 074501.
     [\href{http://arxiv.org/abs/0912.4186}{\tt arXiv:0912.4186 [hep-th]}

\bibitem{Ilgenfritz:2002qs}
     E.-M.~Ilgenfritz, B.~V.~Martemyanov, M.~M\"uller-Preussker, S.~Shcheredin, 
     and A.~I.~Veselov, 
     Phys.\ Rev.\ {\bf D66} (2002) 074503.
     [\href{http://arxiv.org/abs/hep-lat/0206004}{\tt arXiv:hep-lat/0206004}]]

\bibitem{Ilgenfritz:2006ju}
     E.--M.~Ilgenfritz, B.~V.~Martemyanov, M.~M\"uller-Preussker, and A.~I.~Veselov,
     Phys.\ Rev.\ {\bf D73} (2006) 094509.
     [\href{http://arxiv.org/abs/hep-lat/0602002}{\tt arXiv:hep-lat/0602002}]]

\bibitem{Bornyakov:2007fm}
     V.~G.~Bornyakov, E.--M.~Ilgenfritz, B.~V.~Martemyanov, S.~M.~Morozov, M.~M\"uller-Preussker, and A.~I.~Veselov,
     Phys.\ Rev.\ {\bf D76} (2007) 054505.
     [\href{http://arxiv.org/abs/0706.4206}{\tt arXiv:0706.4206 [hep-lat]}

\bibitem{Bornyakov:2008im}
     V.~G.~Bornyakov, E.--M.~Ilgenfritz, B.~V.~Martemyanov, and M.~M\"uller-Preussker,
     Phys.\ Rev.\ {\bf D79} (2009) 034506.
     [\href{http://arxiv.org/abs/0809.2142}{\tt arXiv:0809.2142 [hep-lat]}

\bibitem{GonzalezArroyo:1995ex}
     A.~Gonzalez-Arroyo, P.~Martinez, and A.~Montero,
     Phys.\ Lett.\ {\bf B359} (1995) 159.
     [\href{http://arxiv.org/abs/hep-lat/9507006}{\tt arXiv:hep-lat/9507006}]]

\bibitem{GonzalezArroyo:1996jp}
     A.~Gonzalez-Arroyo and A.~Montero,
     Phys.\ Lett.\ {\bf B387} (1996) 823.
     [\href{http://arxiv.org/abs/hep-th/9604017}{\tt arXiv:hep-th/9604017}]]

\bibitem{GonzalezArroyo:1996gs}
     A.~Gonzalez-Arroyo and A.~Montero,
     Nucl.\ Phys.\ Proc.\ Suppl.\ {\bf 53} (1997) 596.
     [\href{http://arxiv.org/abs/hep-lat/9608035}{\tt arXiv:hep-lat/9608035}]]

\bibitem{GarciaPerez:1989gt}
     M.~Garcia Perez, A.~Gonzalez-Arroyo, and B.~Soderberg,
     Phys.\ Lett.\ {\bf B235} (1990) 117.

\bibitem{GarciaPerez:1992fj}
     M.~Garcia Perez and A.~Gonzalez-Arroyo,
     J.\ Phys.\ {\bf A26} (1993) 2667.
     [\href{http://arxiv.org/abs/hep-lat/9206016}{\tt arXiv:hep-lat/9206016}]]

\bibitem{GarciaPerez:1993ab}
     M.~Garcia Perez, and others,
     Phys.\ Lett.\ {\bf B305} (1993) 266.
     [\href{http://arxiv.org/abs/hep-lat/9302007}{\tt arXiv:hep-lat/9302007}]]

\bibitem{Greensite:2003bk}
     J.~Greensite,
     Prog.\ Part.\ Nucl.\ Phys.\ {\bf 51} (2003) 1.
     [\href{http://arxiv.org/abs/hep-lat/0301023}{\tt arXiv:hep-lat/0301023}]]

\bibitem{Alkofer:2006fu}
     R.~Alkofer and J.~Greensite,
     J.\ Phys.\ {\bf G34} (2007) S3.
     [\href{http://arxiv.org/abs/hep-ph/0610365}{\tt arXiv:hep-ph/0610365}]]

\bibitem{Zakharov:2006vt}
     V.~I.~Zakharov,
     [\href{http://arxiv.org/abs/hep-ph/0602141}{\tt arXiv:hep-ph/0602141}]]

\bibitem{GonzalezArroyo:1998ez}
     A.~Gonzalez-Arroyo and A.~Montero,
     Phys.\ Lett.\ {\bf B442} (1998) 273.
     [\href{http://arxiv.org/abs/hep-th/9809037}{\tt arXiv:hep-th/9809037}]]

\bibitem{Langfeld:2009es}
     K.~Langfeld,
     PoS (QCD-TNT09) (2009) 022.
     [\href{http://arxiv.org/abs/0911.0319}{\tt arXiv:0911.0319 [hep-lat]}

\bibitem{Huang:1996wj}
     S.~z.~Huang,
     Phys.\ Rev.\  {\bf D 54} (1996) 5280.
     [\href{http://arxiv.org/abs/hep-ph/9605461}{\tt arXiv:hep-ph/9605461}]]

\bibitem{Langfeld:2007zw}
     K.~Langfeld,
     Phys.\ Rev.\  {\bf D 76} (2007) 094502.
     [\href{http://arxiv.org/abs/0704.2635}{\tt arXiv:0704.2635 [hep-lat]}

\bibitem{Luscher:1995vs}
     M.~Luscher and P.~Weisz,
     Nucl.\ Phys.\  {\bf B 452} (1995) 213.
     [\href{http://arxiv.org/abs/hep-lat/9504006}{\tt arXiv:hep-lat/9504006}]]

\bibitem{Keurentjes:1998uu}
     A.~Keurentjes, A.~Rosly and A.~V.~Smilga,
     Phys.\ Rev.\  D {\bf 58} (1998) 081701
     [\href{http://arxiv.org/abs/hep-th/9805183}{\tt arXiv:hep-th/9805183}]]

\bibitem{Schaden:2004ah}
     M.~Schaden,
     Phys.\ Rev.\  D {\bf 71} (2005) 105012
     [\href{http://arxiv.org/abs/hep-th/0410254}{\tt arXiv:hep-th/0410254}]]

\bibitem{Langfeld:2005mz}
     K.~Langfeld, N.~Lages and H.~Reinhardt,
     PoS {\bf LAT2005}, 201 (2006)
     [\href{http://arxiv.org/abs/hep-lat/0509156}{\tt arXiv:hep-lat/0509156}]]

\end{thebibliography}
\end{document}